\documentstyle[prl,aps,epsfig]{revtex}

\renewcommand{\Im}{{\rm Im}}

\begin{document}
\draft
\widetext

\title{
An $SU(2)$ Formulation of the $t$-$J$ model:  Application to Underdoped Cuprates
}

\author{
Patrick A. Lee$^{1}$ \\
Naoto Nagaosa$^{2}$ \\
Tai-Kai Ng$^{3}$ \\
Xiao-Gang Wen$^{1}$
}

\address{
$^{1}$Department of Physics, MIT, Cambridge, MA 02139 \\
$^{2}$Department of Applied Physics, University of Tokyo, Tokyo 113, Japan \\
$^{3}$Department of Physics, Hong Kong University of Science and Technology,
Clearwater Bay, Hong Kong
}

\date{June, 1995}
\maketitle
\widetext
\begin{abstract}
\leftskip 54.8pt
\rightskip 54.8pt
We develop a slave-boson theory for the $t$-$J$ model at finite doping
which
respect a $SU(2)$ symmetry -- a symmetry previously known to
be important at half filling.
The  mean field phase diagram is found to be consistent
with the phases observed in the cuprate superconductors, which contains
$d$-wave superconductor, spin gap, strange metal, and Fermi liquid
phases. The spin gap phase is best understood as the staggered flux
phase,
which is nevertheless translationally invariant for physical quantities.
The physical electron spectral function shows small Fermi segments
at low doping which continuously evolve into the large Fermi
surface at high doping concentrations. The close relation between 
the $SU(2)$ and the $U(1)$  slave-boson theory is discussed.
The low energy effective theory
for the low lying fluctuations is derived, and new lying modes (which
were over looked in the $U(1)$  theory) are identified. 

\end{abstract}




\section{Introduction}

It is well established that high temperature superconductivity appears in
cuprates when holes are
doped into the parent compound, which is understood to be Mott-Hubbard
antiferromagnetic (AF)
insulators.  Since the parent compound is insulating only by virtue of strong
correlation, it
stands to reason that a strongly correlated model is the requisite starting
point to describe the
cuprates.  The simplest such model is the two-dimensional $t$-$J$ model and a
large effort has been
made to study how the phase diagram evolves from an Heisenberg antiferromagnet
when a concentration
$x$ of holes are introduced.  The doping of a Mott-Hubbard insulator is a new
problem in condensed
matter physics and involves issues quite different from the doping of a band
insulator.  A key
question is the evolution of the Fermi surface with doping.  At low doping, the
unit cell is
doubled in the AF state and the first holes will form small pockets, not unlike
the doping of band
insulators.  The pockets are centered on $(\pi/2,\pi/2)$.\cite{1}
  On the other hand,
when the hole
concentration is large, it is known that a large Fermi surface is formed, with
an area given by
$1-x$, in agreement with Luttinger theorem.\cite{2}
  The point is that the local
moments on the copper are
now counted as part of the conduction electron that makes up the Fermi sea.
The key question is
how this evolution takes place as a function of doping.  It seems quite likely
that the state
for intermediate doping may contain novel features not encountered before.
Indeed, novel concepts
such as quantum spin liquid sates and spin-charge separation were introduced
early on and much work
has gone into the development of a formal theory which exhibits some of these
novel features.\cite{3}  One
line of approach is to start from mean field decoupling \cite{4,5,6,7}
 and study fluctuations
about the mean field
solution, which turns out to be an $U(1)$ gauge theory.\cite{8,9,10}
On the experimental
front, much work has
focused on the underdoped region, defined as the region of hole concentration
between the onset of
superconductivity and the maximal $T_{c}$, because many anomalous properties
are found in the
metallic state in this regime.  For example, unlike optimally doped systems
where the magnetic
susceptibility $\chi$ and the Knight shift are temperature independent,
underdoped cuprates
generally show a reduction in $\chi$ below 400$K$ or so.\cite{11}
At the same time the
specific heat is
found to be suppressed relative to the $T$ linear behavior expected for
conventional metals.\cite{12}  This
behavior suggests the formation of a gap in the excitation spectrum.  This gap
also shows up in the
$c$ axis frequency dependent conductivity,\cite{13}
 but the conductivity in the plane is
not so strongly
affected.  The {in-plane}
DC conductivity 
shows a suppression below about 200$K$ relative to the linear $T$ resistivity
observed at higher temperatures.\cite{14} This suppression can be attributed to 
reduction of the width of the Drude-like peak by a factor of two.
with little effect on the spectral weight.
the reduction of the conductivity is due to the scattering rate rather than to
carrier
concentration.\cite{15}
  These observations suggest that the gap appears only in the
spin, and not the charge
degrees of freedom in the 2D plane and has been loosely referred to as the spin
gap.  We should add
that the strongest gap-like behavior has been seen in the Cu NMR relaxation
rate and in neutron
scattering, both of which are sensitive to spin excitation at momentum $Q =
(\pi, \pi)$.  This latter
phenomenon  usually onsets at a lower temperature of order 200 K and it has
been argued that it
is observed only in bilayer or trilayer materials.\cite{11,16}
We shall take the point of
view that the
behavior at $(\pi, \pi)$ may be a more delicate issue depending on nesting
properties at the Fermi
surface etc. and for the rest of the paper, we shall use the term spin gap to
refer to properties
mentioned earlier which are characteristic of single-layer as well as
multi-layer cuprates.

Very recently, angle resolved photoemission experiments have yielded important
new information
concerning the electronic excitations of underdoped cuprates.  It was
discovered that a gap in
the spectral functions already existed in the normal state.\cite{17,18}
Furthermore, the
size of this gap and
its dependence on $\vec{k}$ space is similar to the $d$-wave
type gap observed in
the superconducting state.  The difference is that in the normal state, the gap
appears to close in
a finite segment near $(\pi/2, \pi/2)$, leaving a ``Fermi surface segment.''
If this energy gap is
related to the spin gap, this observation gives an important boost to the
notion of spin-charge
separation.  This is because when an electron is removed from the plane, as in
photoemission and in
$c$-axis conductivity, one is forced to pay the energy cost
to break the singlet
pairs in the plane,
whereas for in-plane conductivity, charge transport may occur within the spin
singlet sector.  Such a
behavior is in fact a natural consequence of the mean field phase diagram of
the $t$-$J$ model that
has been in existence for some time.\cite{6,7}
  In this theory the constraint of no
double
occupancy is enforced by writing the electron operator
$c_{\alpha i}$ in terms of auxiliary fermions and boson particles
$c_{\alpha i} = f_{\alpha i} b_i^\dagger$
and demanding that each site is occupied by either a fermion or a boson.
In a mean field (MF) treatment, the order parameters
$\chi_{ij} = \langle f_{\alpha i}^\dagger f_{\alpha j}\rangle$ and
$\Delta_{ij}=\langle f_{1i}f_{2j}
-f_{2i}f_{1j}\rangle$
describes the formation of singlets envisioned in Anderson's resonating
valence bond (RVB) picture.\cite{3}
Above the Bose condensation temperature of the bosons, spin charge separation
occurs at the mean
field level.  In particular, in the underdoped regime the fermions are paired
in a $d$-wave state,
leading to a gap in the spin excitation but no gap in the charge excitation.
This scenario has been
used  as  an explanation of the spin gap phenomenon.\cite{19,20}

While the conventional $U(1)$
mean field theory has a lot of attractive features, it
suffers from a
number of defects.  Firstly, when an attempt was made to improve the theory by
including gauge
fluctuations, it was found that the $d$-wave state was unstable.\cite{21}
  Secondly, in
the underdoped
regime, there are indications that the system is unstable to the spontaneous
generation of gauge
fluxes at finite wave vectors.\cite{22}
Such instabilities will lead to a breaking of
translation symmetry
which is not observed experimentally.  We note that it has recently been
suggested that a
modified $d$-wave state with a large gap at the $(0, \pi)$ point and vanishing
gap along a
segment near $(\pi/2, \pi/2)$ may be stable against gauge fluctuations.\cite{23}
However,
the question about finite wave vector instabilities remains.  Such
considerations motivated us to
produce a new formulation of the constraint which generalizes the
a $SU(2)$ theory for half filled $t$-$J$ model to  $t$-$J$ model
away from half filling.\cite{24}
Our hope is that since SU(2) gauge symmetry is an exact symmetry at half
filling, the mean
field approximation of the new formulation may capture more accurately the low
energy degrees of
freedom and may be a better starting point for small $x$.  Indeed, we found
that in the
underdoped region, the mean field solution may be understood as a $d$-wave
pairing state, or
equivalently as a staggered flux (s-flux)
phase, where the gauge flux alternates on even
or odd
sublattices.  These states are related by local SU(2) gauge transformations and
do not break  
translational symmetry.  Furthermore, these states are connected smoothly to
the $\pi$-flux phase
at half filling which has large excitation energy at the $(0, \pi)$ point,
comparable to that at
$(0,0)$.  This is in agreement with photoemission experiments on the insulating
cuprates,
suggesting that the AF state may resemble the $\pi$-flux phase at short
distances.\cite{26,27}  Furthermore, in
the experiment the state at $(0, \pi)$ moves towards the Fermi surface with
doping, which can be
understood in the mean field theory as a gradual closing of the spin gap.  In
this work \cite{24} we also
introduced a residual attraction between the boson and fermions, and show that
this gives to ``Fermi
surface segments'' near the $(\pi/2,  \pi/2)$ point which grows with doping.
Thus, the $SU(2)$ mean
field theory allows us to answer the fundamental question of how the Fermi
surface evolves from hole
pockets near the $(\pi/2, \pi/2)$ point near half filling, to a large Fermi
surface for large doping
concentration.

In this paper we give a more detailed description of the $SU(2)$ theory and we
also offer an
alternative formulation which has some advantage over the original $SU(2)$ mean
field theory,
particularly in the approach to large doping.  More specifically, in the next
section we show that
the $SU(2)$ theory is intimately related to the original $U(1)$ theory.  This leads
us to a new
formulation in terms of a $\sigma$-model of slowly varying boson fields.  This
is discussed in
Sections 3 and 4.  In Section 5 we present detailed calculations of the
electron spectral
function, comparing the original $SU(2)$ mean field approach and the new
$\sigma$-model
formulation.  We also made some modifications of the interaction potential
between fermions and
bosons, which lead to considerable improvement of the spectral function when
compared with
experiments.  In Section 6 we discuss the collective excitations of the theory,
which are $SU(2)$
gauge fields, and we point out the important massless gauge fields in different
parts of the phase
diagram.  In particular, the existence of a massless mode in the staggered flux
phase is an
important new feature of the $SU(2)$ theory compared with the $U(1)$ formulation.
We also briefly
discuss the response to an electromagnetic field of the normal and
superconducting states.

\section{Relation of the $SU(2)$ formulation to $U(1)$ theory}

Affleck $et$ $al$. \cite{25}
 pointed out that the $t$-$J$ model at half filling obeys an
exact $SU(2)$ symmetry.
They introduced the $SU(2)$ doublets
\begin{equation}
\psi_{1i} =
\left( \begin{array}{r}
   f_{1i} \\
   f_{2i}^\dagger
 \end{array} \right) \;,  \;\;\;
\psi_{2i} =
\left( \begin{array}{r}
    f_{2i} \\
   -f_{1i}^\dagger
 \end{array} \right)
\label{defpsi}
\end{equation}
to represent the destruction of a spin up and spin down on site $i$
respectively.  
This expresses the physical idea that a physical up spin can be represented
by an up spin fermion or the absence of  a down spin fermion once the 
constraint is imposed.
The theory is
invariant under the local transformation $\psi_{\alpha i} \rightarrow
g_{i}\psi_{\alpha i}$ where
$g$ is a $2 \times 2$ matrix representation of the $SU(2)$ group.  In the
original formulation, which
we shall refer to as the $U(1)$ theory, this symmetry is broken upon the
introduction of holes.

In Ref. \cite{24} a new formulation of the constraint of no double 
occupation in the
$t$-$J$ model
was introduced which preserves the $SU(2)$ symmetry even away from half filling.
The key step is
the introduction of a doublet of bosons,
\begin{equation}
h_i =
\left( \begin{array}{r}
   b_{1i} \\
   b_{2i}
 \end{array} \right)
\end{equation}
on each site, so that the physical 
electron operator can be written as an $SU(2)$ singlet, $i.e.$
\begin{eqnarray}
c_{1i} &=&  \frac{1}{\sqrt{2}}
h_{i}^{\dagger} \psi_{1i} =\frac{1}{\sqrt{2}}
 \left(  b_{1i}^{\dagger}f_{1i} +
b_{2i}^{\dagger}f_{2i}^{\dagger} \right)  \nonumber \\
c_{2i} &=& \frac{1}{\sqrt{2}}
 h_{i}^{\dagger} \psi_{2i} =\frac{1}{\sqrt{2}}
 \left(
b_{1i}^{\dagger}f_{2i} - b_{2i}^{\dagger}f_{1i}^{\dagger} \right) 
\end{eqnarray}
The $t$-$J$ Hamiltonian
\begin{equation}
H = \sum_{(ij)}  \left[ J  (\vec S_i\cdot \vec S_j - {1\over 4} n_i
n_j) -t(c^\dagger_{\alpha i} c_{\alpha j} + h.c.) \right]
\end{equation}
can now be written in terms of our fermion-boson (FB) fields.
The Hilbert space of the FB system is larger than that of the
$t$-$J$ model.  However, the local $SU(2)$ singlets satisfying
$ ({1\over 2} \psi^\dagger_{\alpha i} \vec\tau \psi_{\alpha i}
+ b^\dagger_i \vec\tau b_i  ) |{\rm phys} \rangle = 0$
form a subspace that is identical to the Hilbert space of the $t$-$J$
model.
On a given site, there are only three states that satisfy the
above constraint.  They are
$f^\dagger_1 |0\rangle$, $f^\dagger_2 |0\rangle$, and
${1\over \sqrt 2}  (b^\dagger_1
+ b^\dagger_2 f^\dagger_2 f^\dagger_1  ) |0\rangle$
corresponding to a spin up and down electron, and a vacancy
respectively.
Furthermore, the FB Hamiltonian,
 as a $SU(2)$ singlet operator, acts within the subspace, and has
same matrix elements as the $t$-$J$ Hamiltonian.  The projection to the
physical subspace is accomplished by introducing a set of three auxiliary
fields
$a_{0i}^\ell$, $\ell$ = 1,2,3, on each site $i$.  The partition function is
written after a standard Hubbard-Stratonovich transformation as
\begin{equation}
Z = \int DhDh^{\dagger}D \psi D \psi^{\dagger}D \vec{a}_{0}DU
e^{-\int_{0}^{\beta} \stackrel{\sim}{L}}
\label{ZSU2}
\end{equation}
where the Lagrangian $\stackrel{\sim}{L}$ is given by
\begin{eqnarray}
\stackrel{\sim}{L} & = & \frac{\tilde J}{2}  \sum_{<ij>} {\rm Tr}
\left[
U_{ij}^{\dagger}U_{ij}
\right]
+  \frac{1}{2}\sum_{i,j,\alpha}
\psi_{\alpha i}^\dagger\left( \partial_{\tau}\delta_{ij}
+  \tilde J U_{ij} \right)
\psi_{\alpha j}  \nonumber \\
& + & \sum_{i\ell} a_{0i}^{\ell}
\left(
\frac{1}{2} \psi_{\alpha i}^{\dagger} \tau^{\ell} \psi_{\alpha i}
+ h_{i}^{\dagger} \tau^{\ell} h_{i}
\right) \nonumber \\
& + & \sum_{ij} h_{i}^{\dagger} \left( (\partial_{\tau} - \mu) \delta_{ij} +
\tilde t U_{ij} \right)
h_{j}
\label{LSU2}
\end{eqnarray}
The matrix
\begin{equation}
U_{ij} =
\left[ \begin{array}{cc}
   -\chi_{ij}^{*} & \Delta_{ij} \\
   \Delta_{ij}^{*} & \chi_{ij}
 \end{array} \right]
\end{equation}
where $\chi_{ij}$ represents fermion hopping and $\Delta_{ij}$ represents
fermion pairing, respectively and $\tilde J=3J/8$, $\tilde t=t/2$.\cite{28}
The density of physical holes equals the total
density of bosons

\begin{equation}
<1 - c_{\alpha i}^{\dagger}c_{\alpha i}> = <h_{i}^{\dagger}h_{i}> =
<b_{1}^{\dagger}b_{1} +
b_{2}^{\dagger}b_{2}> = x
\label{cbx}
\end{equation}
and is enforced by the chemical potential $\mu$.

The $a_{0i}^{\ell}$ enforces the local constraint
\begin{equation}
<\frac{1}{2} \psi_{\alpha i}^{\dagger} \tau^{\ell} \psi_{\alpha i}
+ h_{i}^{\dagger} \tau^{\ell} h_{i}> = 0
\label{cn}
\end{equation}
In particular, for $\ell = 3$ we have
\begin{equation}
<f_{\alpha i}^{\dagger}f_{\alpha i} + b_{1i}^{\dagger}b_{1i} -
b_{2i}^{\dagger}b_{2i}> = 1
\label{cn3}
\end{equation}

The Lagrangian is invariant under the local $SU(2)$ transformation
\begin{eqnarray}
\psi_{\alpha i} & \rightarrow & g_{i}^\dagger \psi_{\alpha i} \nonumber \\
h_{i}           & \rightarrow & g_{i}^\dagger h_{i}            \nonumber \\
U_{ij}          & \rightarrow & g_{i}^\dagger U_{ij}g_{j}\nonumber \\
a_{0i}^\ell \tau^\ell  & \rightarrow & g_{i}^\dagger a_{0i}^\ell \tau^\ell
g_{i} - 
 g_{i}\partial_\tau g_{i}^\dagger
\label{Gtran}
\end{eqnarray}
where $g_{i}(\tau)$ is a $2 \times 2$ matrix which represents an $SU(2)$ group
element.

Equation \ref{ZSU2} and \ref{LSU2} 
is a faithful representation of the $t$-$J$ model, just as the
more standard $U(1)$ representation is.  The two representations must be
equivalent,  as long as we include all the fluctuations.  
To understand the relation between the $SU(2)$ and the $U(1)$
theory, we will rewrite the $SU(2)$ theory to make it as similar
to the $U(1)$ theory as possible.
In Appendix A we will do the reverse, {\it i.e.} we will
start with the $U(1)$ theory and write it in the  form
of the $SU(2)$ theory, we will also discuss some subtleties of the relation.

The key ingredient is that the two component boson field in the $SU(2)$
representation
is nothing but an $SU(2)$ rotation of the standard slave boson $b_{i}$, $i.e.$
\begin{equation}
h_i = g_{i}
\left( \begin{array}{r}
   b_{i} \\
   0
\end{array} \right)
\label{hgb}
\end{equation}

The matrix $g_{i}$ can be parametrized as
\begin{equation}
g_{i} =
\left( \begin{array}{rr}
   z_{i1} & -z_{i2}^* \\
   z_{i2} & z_{i1}^*
\end{array} \right)
\label{gz}
\end{equation}
with the constraint $\sum_\alpha z_{i\alpha}^*z_{i\alpha} = 1$, which is
satisfied by the parameterization
\begin{eqnarray}
z_{i1} & = & e^{-i \frac{\alpha}{2}} e^{-i\frac{\phi}{2}} \cos \frac{\theta}{2}
\nonumber \\
z_{i2} & = & e^{-i\frac{\alpha}{2}} e^{i \frac{\phi}{2}} \sin
\frac{\theta}{2}
\label{zphi}
\end{eqnarray}
It is natural to introduce the iso-spin vector $\vec{I}$
\begin{eqnarray}
\vec{I} & = & z_\alpha^{*}
\vec{\tau}_{\alpha\beta}z_\beta \nonumber \\
         & = & \left( \sin \theta \cos \phi , \sin \theta \sin \phi , \cos
 \theta \right)
\label{Iz}
\end{eqnarray}
Furthermore it is easy to check that
\begin{equation}
g_{i}\tau_{3}g_{i}^{\dagger} = \vec{\tau} \cdot
\vec{I} .
\label{gI}
\end{equation}
Thus $\vec{I}$ has the meaning of the local quantization axis
parametrized by the polar coordinates $\theta$ and $\phi$.  The angle $\alpha$
in $z_{i}$ and $g_{i}$ is redundant and can be absorbed into the phase of
$b_{i}$
in Eq. \ref{hgb}.  Using Eq. \ref{hgb} 
we can write Eq. \ref{ZSU2} and \ref{LSU2} as
\begin{equation}
Z = \int
DgDbDb^{\dagger}D\psi D\psi^{\dagger}
D\vec{a}_{0}DUe^{-\int_{0}^\beta
L^\prime}
\label{ZSU2a}
\end{equation}
where
\begin{eqnarray}
L^\prime & = & 
\frac{\tilde J}{2} \sum_{<ij>} {\rm Tr}(U_{ij}^\dagger U_{ij}) +
\frac{1}{2}\sum_{ij\alpha}  \psi_{\alpha i}^{\dagger} \left[
    (\partial_{\tau} +\vec a_{0}\cdot \vec \tau) \delta_{ij} + \tilde J U_{ij}
\right]
                          \psi_{\alpha j} \nonumber \\
& + & \sum_{ij} b_{i}^{*} \left( \partial_{\tau} - \mu + \frac {1}{2} {\rm Tr}
\left[ \tau_{3} \left( g_{i}^{\dagger}\vec{a}_{0}\cdot\vec{\tau} g_{i} 
- \left( \partial_{\tau}g_{i}^{\dagger} \right) g_{i} \right) \right]\delta_{ij}
  +  \frac{\tilde t}{2} {\rm Tr}
 \left[ (1 + \tau_{3}) g_{i}^{\dagger}
                            U_{ij}g_{j}\right] \right)
                            b_{j}
\label{LSU2a}
\end{eqnarray}
We see that the path integral of the $SU(2)$ theory is very similar
to that of the $U(1)$ theory.
Here note that the first line of Eq. \ref{LSU2a} is
invariant against the local gauge transformation Eq.11.
To see this we transform of the integral variables
$\psi_{\alpha i} = g_i^\dagger \tilde{\psi}_{\alpha i}$, 
$ g_i^\dagger U_{ij} g_j = \tilde{U}_{ij}$ and 
$
 g_{i}^{\dagger}\vec{a}_{0}\cdot\vec{\tau} g_{i} 
- \left( \partial_{\tau}g_{i}^{\dagger} \right) g_{i}  = 
\vec{ \tilde{a}}_{0}\cdot\vec{\tau} 
$. 
Then Eq. \ref{ZSU2a} and \ref{LSU2a} become
\begin{equation}
Z = \int
DgDbDb^{\dagger}D\tilde\psi D\tilde\psi^{\dagger}
D\vec{\tilde a}_{0}D\tilde Ue^{-\int_{0}^\beta
\tilde L^\prime}
\label{ZSU2a1}
\end{equation}
and
\begin{eqnarray}
\tilde L' &=& { \tilde J \over 2} \sum_{<ij>} {\rm Tr} 
\biggl[ \tilde U^{\dagger}_{ij}\tilde U_{ij} \biggr]
+ \frac{1}{2} \sum_{ij\alpha} \tilde \psi_{i\alpha}^{\dagger}
\biggl[ (\partial_\tau + \sum_{a=1}^3 \tilde a^{a}_0 \tau_a )
\delta_{ij}  + \tilde J\tilde U_{ij} \biggr] \tilde \psi_{j\alpha}
\nonumber \\
&+&   \sum_i b^{\dagger}_i ( \partial_\tau -  \mu + \tilde a^3_0 ) b_i
- \tilde t \sum_{ij} \chi_{ij} b^{\dagger}_j b_i
\label{LSU2a1}
\end{eqnarray}
Note that $\tilde L'$ no longer depends on $g$ so that the $g$ integral can
be dropped. If we drop $a_0^{1,2}$ integral, Eq. \ref{ZSU2a1} and
\ref{LSU2a1} have the same form as the $U(1)$ formulation with an exception
that there $\tilde t$ is replaced bt $t=2\tilde t$.
It is not our purpose to derive the exact equivalence between the
$U(1)$ and $SU(2)$ path integrals, but rather we want to point out how
low lying fluctuations in the $SU(2)$ formulation may be reproduced
in the $U(1)$ picture.

The $U(1)$ mean field theory corresponds to fixing $g$ to be unity (so that
$\vec{I} = \hat{z}$) and finding $U_{ij}^{0}$ and
$\vec{a}_0^{(0)}$ which minimizes the action after summing
over
$\psi$ and $b$.  In the underdoped region, it was found that $U_{ij}^{(0)}$
corresponds to $d$-wave pairing of fermions.  Thus the $SU(2)$ symmetry at half
filling
is broken by the boson term for finite $x$.  At the same time, it is clear that
for $x \ll 1$, there is a host of $U(1)$ mean field states $U_{ij} =
g_{i}^{\dagger}U_{ij}^{(0)}g_{j}$ which are close in energy to the $d$-wave
state.
Since these states are degenerate at $x = 0$, we may expect an energy cost of
order $xJ$ per hole, or $x^{2}J$ per unit cell.  An example of special interest
is the staggered flux phase which has a Dirac spectrum $E_{k} =
\sqrt{\xi_{k}^{2}
+ \Delta_{k}^{2}}$ at $\left( \pi/2 , \pi/2 \right)$.  Since the density of
states of the Dirac spectrum is linear in energy, the energy cost is $\sim
\mu_{F}^{3}/\Delta J$ for a given fermion chemical potential. To satisfy the
fermion number
constraint, $\mu_{F}  \approx \sqrt{x\Delta J}$ so that in this case the energy
cost is expected to
be $\sqrt{\Delta J}$ $x^{3/2}$ per unit cell.  
At finite temperatures, we expect that these low
energy configurations should be included in the partition function sum.  This
additional degree of freedom is just represented by the functional integral
over
$g$ in Eq. \ref{ZSU2a}, 
and this is the motivation for adopting the $SU(2)$ formulation.

In Ref. \cite{24}
 a mean field theory was introduced for the $SU(2)$ action Eq. \ref{ZSU2}
and \ref{LSU2}.  The
mean field is a saddle point of the action with respect to $U_{ij}$ and
$\vec{a}_{0}$, after integrating over $\psi,\psi^{\dagger}$
and
$h$,$h^{\dagger}$, which is possible because the action is quadratic in these
variables.  We find that the mean field phase diagram is only slightly modified
from the $U(1)$ case, and consists of six different phases.

\noindent
{(1)} Staggered flux (s-flux) phase:
\begin{eqnarray}
U_{i, i + \hat x} & = &- \tau^3 \chi - i (-)^{i_x+i_y} \Delta \nonumber\\
U_{i, i + \hat y} & = &- \tau^3 \chi + i (-)^{i_x+i_y} \Delta
\label{Uij}
\end{eqnarray}

\noindent
and $a^l_{0i}  = 0$.
In the $U(1)$ slave-boson theory, the staggered flux phase breaks
translational symmetry.
Here the breaking of translational invariance is a gauge
artifact.
In fact, a site dependent $SU(2)$ transformation
$W_i = {\rm exp} [ i (-1)^{i_x + i_y} (\pi/4) \tau_1 ]$
 maps the s-flux
phase to the $d$-wave pairing phase of the fermions:
$U_{i, i+ \hat x,\hat y} = -\chi\tau_3 \pm \Delta\tau_1$,
which is explicitly translationally invariant.
In the s-flux phase
 the fermion and boson dispersion are given by
$\pm E_f$ and $\pm E_b$, where $E_f =
 \sqrt{(\epsilon_f-a_0^3)^2 + \eta^{2}_f}$, $\epsilon_f  =
-2\tilde J (\cos k_x + \cos k_y ) \chi$,
$\eta_f  = -2\tilde J (\cos k_x - \cos k_y ) \Delta$,
and a similar result for $E_b$ with $\tilde J$ replaced by $\tilde t$.
Since $ia^3_0 = 0$ we have
$\langle f^\dagger_{\alpha i} f_{\alpha i} \rangle = 1$ and
$\langle b^\dagger_1b_1\rangle =
\langle b^\dagger_2b_2\rangle = x/2$.

\noindent
{(2)}  The $\pi$-flux ($\pi F$) phase is the same as the s-flux phase
except
here $\chi = \Delta$.

\noindent
{(3)}  The uniform RVB (uRVB) phase is described by Eq. \ref{Uij}
with $a_{0i}^l=\Delta = 0$.

\noindent
{(4)}  A localized spin (LS) phase has
$U_{ij} = 0$ and $a^l_{0i} = 0$, where
the fermions cannot hop.

\noindent
{(5)}  The $d$-wave superconducting (SC) phase is described by
$U_{i, i+ \hat x,\hat y} = -\chi\tau_3 \pm \Delta\tau_1$
and $a^3_0 \ne 0$, $a^{1, 2}_0 = 0$, $\langle b_1\rangle \ne 0$,
$\langle b_2\rangle = 0$. 

\noindent
{(6)}  The Fermi liquid (FL) phase is similar to the SC
phase except that there is no fermion pairing ($\Delta = 0$).

The connection with the $U(1)$ mean field theory is now clear by using 
Eq. \ref{LSU2a}.
The $SU(2)$ mean field consists of fixing $U_{ij} = U_{ij}^{(0)}$ and
$\vec{a}_{0} = \vec{a}_{0}^{(0)}$.
For each $\{g_i\}$ the integral over $\psi,\psi^\dagger$, $b,b^\dagger$ gives
the free energy of a $U(1)$ mean field theory with
\begin{equation}
U_{ij}(g) = g_i^{\dagger}U_{ij}^{(0)}g_j
\label{Ug}
\end{equation}
and
\begin{equation}
\vec{a}_0 \cdot \vec{\tau} =
g_i^{\dagger}a_0^{(0)} \cdot \vec{\tau} g_i +
g_i^{\dagger} \partial_\tau  g_i
\label{a0g}
\end{equation}
Upon integration over $\{g_i\}$, we see that the $SU(2)$ mean field theory
includes
the $U(1)$ mean field state $\{U_{ij}^{(0)},\vec{a}_0^{(0)}\}$ and
all the
configurations
$\{U_{ij},\vec{a}_0\}$ 
 connected to it by $SU(2)$ rotations.  Thus for $x \ll 1$ all the
low energy excitations are included in the partition sum.  This is the reason
why
we believe the $SU(2)$ mean field theory is a better starting point for
underdoped
cuprates.

We note that with the exception of the superconducting and Fermi liquid phases,
$\vec{a}_0^{(0)} = 0$ in the $SU(2)$ 
mean field solution.  This
means that $\langle f_{\alpha i}^\dagger f_{\alpha i} \rangle = 1$ and the
constraint Eq.
\ref{cn3} is satisfied by $\langle b_1^{\dagger}b_1 \rangle - \langle
b_2^{\dagger}b_2 \rangle =
0$.  Unlike the $U(1)$ case, the density of fermions is not necessarily $1-x$.
It is
this feature which allows the staggered-flux and $d$-wave states to be gauge
equivalent descriptions in the s-flux phase, for instance.  One consequence is that
the node in the gap function of the fermion excitation is pinned at $(\pi
/2,\pi
/2)$.  In Ref. \cite{24}
 it was found that by including an attraction between the
boson and fermion due to the exchange of $a_0$ fluctuations, Fermi surface-like
features can be recovered in the physical electron spectral weight which is
shifted away from $(\pi/2, \pi/2)$.  

A similar situation appears in the uRVB phase.  The fermion Fermi surface
encloses area 1 and one must go beyond mean field theory to produce an electron
Fermi surface-like features which obey the Luttinger theorem.  The problem is
even
more serious in the FL phase.  Even though $a_0^3$ is now not equal to zero, the
fermion Fermi surface area approaches $1 - x$ only very slowly with increasing
$x$ and
decreasing temperature.  Granted that the FL state exists only for $x \geq J/t$
so
that the motivations behind the $SU(2)$ mean field theory is no longer
applicable.
Nevertheless, this observation means that the $SU(2)$ mean field theory does not
evolve towards the $U(1)$ mean field theory in a way which is acceptable.

We believe the origin of these difficulties lies in fixing
$\vec{a}_0^{(0)}$ as a
mean field parameter from the beginning.  For
$\vec{a}_0^{(0)} = 0$,
the constraint is satisfied on the average by $\langle h^\dagger
\vec{\tau} h \rangle = 0$.  For example, this implies
$\langle b_{1i}^{\dagger}b_{2i} \rangle = 0$.  Using Eq. \ref{gz} and 
\ref{Iz}, this
suggests
that the iso-spin vector $\vec{I}$ is randomized so that
$\langle \vec{I}_i \rangle = 0$.  On the other hand, as we
approach the superconducting phase boundary $T_c$ from above, or the Fermi
liquid
boundary from the uRVB side, the boson field $h_i$ is becoming phase coherent
and
we expect that it should be slowly varying in space and time.  In these
regions,
the short range correlation of the boson field is not captured by the $SU(2)$
mean
field theory.  This motivates us to formulate a new effective theory for the
$SU(2)$ partition function which we shall refer to as the $\sigma$-model
description.

Our strategy is to pick a mean field configuration $U_{ij}^{(0)}$ and consider
a slowly varying
configuration $h_{i}$ in Eq. \ref{LSU2}
 or, equivalently, a slowly varying $g_{i}$ and
$b$ in Eq. \ref{LSU2a}.
For each configuration, $\vec{a}_{0}$ is solved to satisfy
the constraint
locally, after performing the integral over $\psi$, $\psi^\dagger$.  Thus in
principle
$\vec{a}_{0}$ is a functional of $\{
\vec{h}_{i} \}$.  Our final
goal is to produce an effective Lagrangian for $\{
\vec{h}_{i} \}$ which will take
the form of some nonlinear $\sigma$-model to describe the low energy physics of
the problem.  This
is the opposite limit to the $SU(2)$ mean field theory: the assumption of a
uniform
$\vec{a}_{0}$ is valid when the {$h_{i}$} configurations are
rapidly varying on
the scale of the fermion correlation length, which is of order $\xi_0 =
\epsilon_{F}/\Delta$ in the
s-flux phase.  This picture is valid at high temperatures, whereas the
$\sigma$-model approach is
expected to be applicable near the superconducting transition and the crossover
to the Fermi liquid
state.  The truth most likely lies in between the two extreme limits in most
parts of the phase
diagram, and it will be of interest to explore the consequences of both limits.

It is clear that any $\tilde{U}_{ij}^{(0)}$ related to $U_{ij}^{(0)}$ by a
$SU(2)$ gauge
transformation will give an equivalent description.  Thus we can start with any
$U(1)$ mean field
configuration.  In principle, we should optimize the parameters $\chi$ and
$\Delta$ at the end of the
calculation, but in practice we expect these parameters to be not so different
from that given by
the $U(1)$ mean field theory.  We also find that a judicial choice of
$U_{ij}^{(0)}$ which exhibits
the symmetry of a given phase yields a $\sigma$-model which exhibits the proper
symmetry.  As a
first example we discuss the uRVB state.

\section{$\sigma$-model of the Fermi liquid and the uRVB phases}

In $U(1)$ mean field theory the matrix $U_{ij}^{(0)}$ in the uRVB state
is given by
$
U_{ij}^{(0)} =
\left( \begin{array}{cc}
 -\chi_{ij}^{*} & 0 \\
   0            & \chi_{ij}
\end{array} \right)
$.
Here we make the choice $\chi_{ij} = i\chi_{0}$ so that $U_{ij}^{(0)} =
i\chi_{0}I$ is proportional
to the identity element.  Thus $U_{ij}^{(0)}$ itself is invariant under a
global $SU(2)$
transformation.

For $a_{0}^{1} = a_{0}^{2} = 0$ the bosons $b_{1}$ and $b_{2}$ are diagonalized
by the energy
dispersion
\begin{equation}
E_{b}^{1,2} = - 2t \chi_{0} \left( \sin k_{x} + \sin k_{y} \right) \pm 
a_{0}^{3} - \mu
\end{equation}
In the Fermi liquid phase, the boson condenses to the bottom of the band,
located for this choice
of gauge at $Q_{0} = (\pi /2, \pi /2)$.

As explained in Ref. \cite{24}
the $SU(2)$ mean field theory solution for the Fermi liquid
is given by $a_{0}^{3}
< 0$, and $b_{1}$ contains a Bose condensed part so that $\langle b_{1} \rangle
= b_{0} e^{i
\vec{Q}_{i} \cdot \vec{r}}$.  Note that at
finite $T$,
thermal excitations make $\langle b_{2}^\dagger b_{2} \rangle \neq 0$.  From
Eq. \ref{cbx} and \ref{cn3} we
see that the fermion density
\begin{equation}
\langle
\sum_\alpha f_{i\alpha}^\dagger f_{i\alpha}
\rangle
= 1 - x + 2
\langle
b_{2}^\dagger b_{2}
\rangle
\end{equation}
is not equal to $1 - x$ so that Luttinger theorem is not obeyed.  As
discussed in the introduction, this
motivates us to try the
$\sigma$-model approach, where we write
\begin{equation}
h_{i} = \tilde{h}_{i} e^{ i\vec{Q}_{0} \cdot
\vec{r} }
\label{locU1}
\end{equation}
and look for $\tilde{h}_{i}$ which is slowly varying in space and $\tau$.  
We can further parameterize $\tilde{h}_{i} = g_{i}
\left(
\begin{array}{r}
b \\
0
\end{array}
\right)$.
Locally we can consider $g_{i} = g_{0}$ as constant.  By introducing
$\tilde{\psi} = g_{i}^\dagger
\psi$ we see that $L^\prime$ in Eq. \ref{LSU2a} takes the $U(1)$ form
\begin{eqnarray}
L^\prime & = & \frac{1}{2}
\sum_{ij\alpha}  \tilde{\psi}_{i\alpha}^\dagger
\left[
(\partial_{\tau} + \vec{a}_{0}^\prime \cdot
\vec{\tau})
  \delta_{ij} + \tilde J U_{ij}^{(0)}
\right]
\tilde{\psi}_{i\alpha}  \nonumber \\
& + & \sum_{ij} b^{*}_i
\left(
\partial_{\tau} - \mu + \frac{1}{2} {\rm Tr}  \left( \tau_{3}
(\vec{a}'_0
\cdot \vec{\tau}) \right) \delta_{ij}
 + \frac{\tilde{t}}{2} {\rm Tr} \left[ (1 + \tau_{3}) U_{ij}^{(0)} \right]
\right)
b_{j}
\label{LuRVB0}
\end{eqnarray}
where
\begin{equation}
\vec{a}_{0}^\prime \cdot \vec{\tau} =
g_{i}^\dagger \vec{a}_{0} \cdot \vec{\tau}
g_{i} - \left(
\partial_{\tau}g^{\dagger}\right) g
\end{equation}

The local $U(1)$ mean field solution of Eq. \ref{locU1} is given by
$\vec{a}_{0}^\prime =
a_{00} \hat{z}$ and $a_{00}$ is the Fermion chemical potential chosen in a way
which ensures that
the $\tilde{\psi}$ fermion density is $1 - x$.
{}From Eqs. \ref{locU1} and \ref{gI}, we find that
\begin{equation}
\vec{a}_{0i} = a_{00} \vec{I}(g_{i})
\end{equation}
The physical electron Green's function in the $SU(2)$ theory
\begin{equation}
G(\vec{r},\tau)  =  -\langle T_\tau \left(
c_1 (\vec{r},\tau) c_1^\dagger(
\vec{0},0) \right)
\rangle 
 =  -\frac{1}{2}
\langle T_\tau \left( (h^\dagger (\vec{r},\tau)
\psi_1
(\vec{r},\tau)
\psi_1^\dagger (\vec {0},0) h (
\vec{0},0) \right)
\rangle  
\end{equation}
Assuming $b$ is Bose condensed, we have, within the mean field theory,
\begin{equation}
G(r,\tau) = -\frac{1}{2}
b_{0}^{2} \langle T_\tau \left(  \tilde{f}_1 (\vec r,\tau)
\tilde{f}_1^\dagger (0,0) \right)  \rangle \;\;\; \rm {+ \;\;\;
incoherent \;\;\; part}
\end{equation}
The fermion part $\tilde{\psi}_{\alpha}$, and therefore the physical electron
Fermi surface  now satisfies Luttinger theorem in
this slowly-varying approximation.

We would like to remark that the electron Green function
in the $U(1)$ theory has a form
\begin{equation}
G(\vec{r},\tau)  =  -\langle T_\tau \left(
c_{\uparrow}(\vec{r},\tau) c_{\uparrow}^\dagger(
\vec{0},0) \right)
\rangle
 =  -
\langle T_\tau \left( (b^\dagger (\vec{r},\tau)
f_{\uparrow}
(\vec{r},\tau)
f_{\uparrow}^\dagger (\vec {0},0) b (
\vec{0},0) \right)
\rangle  
\end{equation}
The $U(1)$ mean field Green function is
\begin{equation}
G(r,\tau) = -
b_{0}^{2} \langle T_\tau \left(  f_{\uparrow} (\vec r,\tau)
f_{\uparrow}^\dagger (0,0) \right)  \rangle +
\hbox{incoherent  part}
\end{equation}
after the boson condensation.
Although the coherent part has the same dispersion relation,
the quasiparticle weight in the $U(1)$ mean field Green function
is twice of the quasiparticle weight in the $SU(2)$ mean field Green function.

We next derive an explicit expression for the $\sigma$-model Lagrangian by
expanding in $a_{0}$ and
integrating out the fermion.  This is a systematic procedure for small $x$.
Starting from
Eq. \ref{ZSU2}, the fermion integration 
yields a contribution $-{\rm Tr} \ln (\partial
_\tau +
\tilde JU_{ij}^{(0)} - i\vec{a}_{0} \cdot
\vec{\tau})$.  An expansion in
$\vec{a}_{0}$ to quadratic order yields the term
\begin{equation}
{\cal L'}_{F} = 
\sum_{q \cdot \omega_n}\frac{1}{2} a_{0}^\alpha
(q,\omega_{n})a_{0}^{*\beta}(q,\omega_{n})\pi_{00}^{\alpha\beta}(q,\omega_{n})
\end{equation}
where
\begin{equation}
\pi_{00}^{\alpha \beta} (q,\omega_{n}) = \int _{0}^\beta d\tau
e^{i\omega_{n}\tau}
\sum_{i}e^{-i\vec{q} \cdot (\vec{r}_{i}-
\vec{r}_{j})} \langle \psi_{i}^\dagger \tau_\alpha \psi_{i}
(\tau)
\psi_{j}^\dagger \tau_\beta \psi_{j}(0) \rangle
\end{equation}
For the uRVB state, $\pi_{00}^{\alpha \beta} = \pi_{00}^{(0)} \delta_{\alpha
\beta}$, where
$\pi_{00}^{(0)}$ may be expanded for small $q$ and $|\omega_{n}| < q$ as
\begin{equation}
\pi_{00}^{(0)} (q,\omega) = \pi_{0} + C_{1}' J^{-1} q^{2} + C_{2}' J^{-2}
\frac{|\omega_{n}|}{q}
\end{equation}
The coefficient $\pi_{0} = -C_{0}'J^{-1}$ 
where $C_{0}'$, $C_{1}'$ and $C_{2}'$ are
constants of order
unity.  The leading term gives a contribution $-C_{0}'J |a_{0}|^{2}$.  The
negative sign is a
reminder that the mean field $a_{0}$ is a saddle point with the stable
direction along the
imaginary axis.  We shall see that this negative sign yields correctly a
repulsive interaction
between the bosons.

We expand in $\tilde{h}$ about the bottom of the boson bands and the effective
Lagrangian takes the form
\begin{eqnarray}
{\cal L}_{eff} & = & \tilde{h}^\dagger  \partial_{\tau} \tilde{h} +
\frac{1}{2m_{b}}
|\partial_{i}\tilde{h}|^{2} - \mu \tilde h^\dagger \tilde h \nonumber \\
& + & D_{1} m_{b}^{-1} (\tilde{h}^\dagger \tilde{h})^{2} + |b|^{2}
      \vec{a}_{0}(\vec{r},\tau) \cdot
      \vec{I}(\vec{r},\tau) \nonumber \\
& + & \frac{1}{2} \sum_{q,\omega_{n}} |a_{0}(q,\omega_{n})|^{2}
      \pi_{00}^{(0)}(q,\omega_{n})
\label{LuRVB1}
\end{eqnarray}
The $D_{1}$ term is used to model the repulsion between bosons and $D_{1}$ is
of order unity for
infinite on-site repulsion.  We have rewritten 
the coupling between $\tilde{h}$ and
$\vec{a}_{0}$ using Eq. \ref{gI} and \ref{hgb}.  
Since Eq. \ref{LuRVB1} is quadratic in
$a_{0}$, it can be
eliminated, yielding a fermion contribution to the Lagrangian
\begin{eqnarray}
{\cal L''}_{F} = & - \frac{1}{2} & \sum_{q,\omega_{n}} \frac{|b|^{4}
                 \vec{I}^{*}(q,\omega_{n})
                 \cdot
\vec{I}(q,\omega_{n})}{\pi_{00}^{(0)}(q,\omega_{n})}
                 \nonumber \\
               & \approx & - \frac{1}{2} |b|^{4} \sum_{q,\omega_{n}}
                     \vec{I}^{*}(q,\omega_{n}) \cdot
\vec I(q,\omega_{n})
                      \left( - C_{0} J + C_{1} Jq^{2} + C_{2}
\frac{|\omega_{n}|}{q} \right)
\end{eqnarray}
Using $\vec{I} \cdot \vec{I} \equiv 1$, the
first term is
$\frac{1}{2} C_{0}J|b|^{4}$ and it modifies the $D_{1}$ term in Eq. \ref{LuRVB1}
 to $D_{1}^\prime =
D_{1}+C_{0}J$.  For $J < t$, this is a small correction.  To obtain a
description in terms of
the $z$ fields $z = (z_{1},z_{2})$ where $z_{1},z_{2}$ are defined in 
Eq. \ref{gz},
we write
$\tilde{h} = (b_{0} + \delta b)z$ and integrate out the $\delta b$ field.  We
find
\begin{eqnarray}
{\cal L}_{eff} & = & \frac{2}{3} \frac{m_{b}}{D_{1}^\prime} |z^\dagger \partial
_\tau z|^{2} +
                     |b_{0}|^{2} z^\dagger \partial _\tau z + \frac{x}{2m_{b}}
|\partial_{i}z|^{2}
                     + {\cal L}_{F} 
\label{LuRVBb}\\
{\cal L}_{F}   & = & \frac{1}{2} x^{2}J \sum_{q,\omega_{n}}
\vec{I}^{*}
                     (q,\omega_{n}) \cdot
\vec{I}(q,\omega_{n})
                     \left(  C_{1}q^{2} + C_{2} \frac{|\omega_{n}|}{q} \right)
\label{LuRVBf}
\end{eqnarray}
We have approximated $b^{2}$ by $x$ and $b_{0}$ is a constant of order
$\sqrt{x}$ at low
temperature.  The first term in Eq. \ref{LuRVBf} is a ferromagnetic Heisenberg
interaction between the
isospins.  Using the usual CP$^1$ representation, it can be written as
\begin{equation}
\frac{C_1}{2} x^{2}J| (\partial_{i} - i\tilde{A}_{i})z|^{2}
\end{equation}
where
\begin{equation}
\tilde{A}_{i} = \frac{i}{2} (z^{\dagger}\partial_{i}z 
- (\partial_{i}z^{\dagger}) z)
\end{equation}
Note that whereas the boson part  in Eq. \ref{LuRVBb}
 has the full O(4) symmetry, the
fermion part only has
O(3) symmetry { because it is independent of the overall phase} $\alpha$.
The second term in Eq.  \ref{LuRVBf}
 describes dissipation due to particle-hole excitations of the Fermi sea.
Note that the fermion
contribution is proportional to $x^{2}J$ which is smaller than the boson
contribution which is
proportional to $xt$ even in the overdoped region $(xt \geq J)$.  For example,
if $T > x^{2}J$ we
can ignore the fermion term and if we further make the classical approximation,
we conclude that at
high temperatures the system is described by the classical O(4) model.  There
is no phase
transition but instead there is a cross-over temperature of order $xt$ below
which the phase
coherence length grows exponentially.  This is opposed to the $U(1)$ mean field
theory where there
is a Kosterlitz-Thouless transition.  Of course this transition is destroyed
when gauge
fluctuations are taken into account.\cite{29}
  However, in our case  vortex excitations
are destroyed by $SU(2)$
fluctuations and we can expect a suppression of the development of phase
coherence in the $SU(2)$
formulation due to the addition of low lying degrees of freedom.  It is
interesting to ask, what is
the nature of these low energy excitations?  In the effective Lagrangian Eq.
\ref{LuRVBb}, the degeneracy for
constant $z$ is a gauge symmetry: any constant $z$ is related by a global gauge
transformation to the $U(1)$ uRVB state.  When $z$ is slowly varying, we can use
Eq. \ref{Ug} to see that
in the $U(1)$ representation, $U_{ij} = i\chi_{0}g_{i}g_{j}^\dagger$ are
generated which in general
contains pairing amplitudes $\Delta_{ij}$ as well as modifications of the
hopping term $\chi_{ij}$
which affects both the boson and fermion energy.  
This is in contrast to  the $U(1)$ formulation, where only the phase
fluctuation of $\chi_{ij}$ is included.  
Thus we may view the $SU(2)$ formulation
as a way to discover
low lying excitations which were not so obvious in the $U(1)$ picture.  To
complete the discussion of
the low lying excitations we need to introduce gauge fields to the effective
Lagrangian.  This will
be done in a later section.

\section{$\sigma$-model of the superconducting and the staggered flux phases}

We repeat the procedure in the last section for the staggered flux phase by
choosing an
appropriate $U_{ij}^{(0)}$ matrix.  Once again any $U_{ij}^{(0)}$ which are
related by gauge
transformations will give the same result, but it will be convenient to use
a $U_{ij}^{(0)}$ which
exhibits the symmetry of the state.  We have noted before that in the $SU(2)$
mean field theory, the
s-flux state breaks the $SU(2)$ symmetry down to $U(1)$.  This motivates us to
choose the following
$U_{ij}^{(0)}$ to describe the s-flux phase.

We choose the following ansatz to describe the s-flux phase
\begin{equation}
U_{i,i+\hat x}^{(0)} = -i\chi -(-)^i \tau^3 \Delta,\ \ \ \ \
U_{i,i+\hat y}^{(0)} = -i\chi +(-)^i \tau^3 \Delta
\label{a2}
\end{equation}
and
\begin{equation}
a_0^l(i)=a_0^l+(-)^i \tilde a_0
\end{equation}
Note that $U_{ij}^{(0)}$ is invariant under global $\tau_{3}$ rotations.
In the momentum space $\psi_i=\sum_i e^{-i\vec k \cdot \vec i} \psi_k$,
we have
\begin{eqnarray}
H_{mean}^f   &=& \tilde J {\sum_k}^\prime
\pmatrix{\psi_k^\dagger ,& \psi_{k+Q}^\dagger \cr}
\pmatrix{ V_k+a_0^l\tau^l,       &  W_{k+Q} +\tilde a_0^l \tau^l \cr
          W_k+\tilde a_0^l\tau^l,&  V_{k+Q} +       a_0^l \tau^l \cr  }
\pmatrix{\psi_k \cr \psi_{k+Q} \cr} \nonumber \\
H_{mean}^b   &=&\tilde  t {\sum_k}^\prime
\pmatrix{h_k^\dagger ,& h_{k+Q}^\dagger \cr}
\pmatrix{ V_k+a_0^l\tau^l,       &  W_{k+Q} +\tilde a_0^l \tau^l \cr
          W_k+\tilde a_0^l\tau^l,&  V_{k+Q} +       a_0^l \tau^l \cr  }
\pmatrix{h_k \cr h_{k+Q} \cr}
\end{eqnarray}
where
\begin{eqnarray}
V_k &= -2\chi(\sin k_x + \sin k_y)       &= -2\chi \alpha_k \nonumber \\
W_k &= -2i\tau^3 \Delta (\sin k_x - \sin k_y) &= -2i\tau^3 \Delta \gamma_k
\label{a3}
\end{eqnarray}
and ${\sum_k}^\prime$ represents summation over half of the Brillouin zone.

To study the boson condensed phase at low temperatures,
let us first assume $\tilde a_0^l=0$.
In this case the boson band bottom is at $k=(\pi/2, \pi/2)$ if
$a_0^l$ are not too large. Thus the condensed boson has a form
\begin{equation}
\pmatrix{b_1(i) \cr b_2(i)}=
\pmatrix{b_1 \cr b_2 \cr} e^{-i(i_x + i_y) \pi/2}
\label{a4}
\end{equation}
For such a boson condensation the boson free energy is an even function
of $\tilde a_0^l$. We also note that
\begin{equation}
M(k_x,k_y,a_0^l,\tilde a_0^l)=
\pmatrix{ V_k+a_0^l\tau^l,       &  W_{k+Q} +\tilde a_0^l \tau^l \cr
          W_k+\tilde a_0^l\tau^l,&  V_{k+Q} +       a_0^l \tau^l \cr  } =
\pmatrix{
  -2 \chi        \alpha_k +       a_0^l \tau^l,
&  2i\Delta\tau^3\gamma_k +\tilde a_0^l \tau^l \cr
  -2i\Delta\tau^3\gamma_k +\tilde a_0^l \tau^l,
&  2 \chi        \alpha_k +       a_0^l \tau^l \cr       }
\label{a5}
\end{equation}
satisfies
\begin{equation}
M(k_x,k_y,a_0^l,\tilde a_0^l)=
\pmatrix{1&&&\cr &1&&\cr &&-1&\cr &&&-1\cr}
M(k_y,k_x,a_0^l,-\tilde a_0^l)
\pmatrix{1&&&\cr &1&&\cr &&-1&\cr &&&-1\cr}
\label{a6}
\end{equation}
Thus the fermion free energy is also an even function of $\tilde a_0^l$.
Therefore $\tilde a_0^l=0$ is a self consistent solution.

We would like to remark that $U_{ij}$ in Eq. \ref{a2} does not contain
any fermion pairing. However the boson condensate induces non-zero
$a_0^l$. A non-zero $a^{1,2}_0$ induces a pairing condensate
of the fermions. But when $a^{1,2}_0=0$ there is no pairing and
the fermions are in a normal Fermi liquid state.

Now we are ready to discuss some basic physical
properties of our ansatz for different orientation of the
condensate $(b_1,b_2)$.
Without lose of the generality, we may assume $b_1/b_2=$real.
In this case $a_0^2=0$.
We see that when $b_1=b_2$ (in this case $a_0^3=0$)
the ansatz describes a translation and rotation invariant state.
This state is equivalent to the usual $d$-wave paired state
in the $U(1)$ meanfield theory after a $SU(2)$ gauge transformation.
It describes a $d$-wave superconducting state (with a finite chemical
potential) of the $t$-$J$ model.
When $b_1\neq b_2$ and $b_1b_2 \neq 0$, we have $a_0^3 \neq 0$ and
$a^1_0\neq 0$. There is a pairing condensate in the fermions.
The ansatz describes
a superconducting state of the $t$-$J$ model which also breaks
the translation symmetry. The quasiparticle excitations have
finite gap except at four isolated points near $(\pm \pi/2, \pm \pi/2)$.
When $b_1\neq 0$ and $b_2=0$, we have $a^3_0 \neq 0$ and $a^{1,2}_0=0$.
There is no pairing condensate in the fermions. The ansatz, despite the boson
condensate, does not correspond to a
superconducting state.   It instead describes
a Fermi liquid with broken translation symmetry and small
pocket-like Fermi surfaces.  This result is obtained
through a later calculation of electromagnetic response.

At high enough temperatures, the thermo fluctuations make
$\langle b_1\rangle  =\langle b_2\rangle =\langle a_0^l\rangle =0$.
In this case the ansatz describes
a translation and rotation invariant metallic state, which is just
the s-flux phase studied in Ref. \cite{24}

In order to derive a $\sigma$-model for the $\tilde{h}$ field, we integrate
out the
fermions as before.  The difference now is that $\pi_{00}^{\alpha \beta} =
\pi_{00}^{\alpha}
\delta_{\alpha \beta}$ where $\pi_{00}^{x} = \pi_{00}^{y} \neq
\pi_{00}^{z}$.  We find that
$\pi_{00}^{x}(0) \approx C_{1}^\prime J$ whereas $\pi_{00}^{z}(0) = 0$ for
$\vec{a}_{0} = 0$.  This is because $\pi^{z}$ is the
density-density response
function and $\pi^{z}(0)$ is the compressibility of the fermion which
vanishes due to the vanishing
density of states in the middle of the band.  For finite $a_{0}^{3}$ we
find that $\pi_{00}^{z}(0)
= C_{2}^\prime a_{0}^{3}$.  Now we can eliminate
$\vec{a}_{0}$ to extremize the
action.  The problem retains rotational symmetry in the $x$-$y$ plane but
is anisotropic in the
$z$ direction.  For example, for $\vec{I}$ in the $x$-$y$
plane, we have
$a_{0}^{3} = 0$ and the energy of the mean field state is
\begin{equation}
E_{MF} = -4t \chi x + \frac{J}{2C_{1}^\prime} x^{2}
\end{equation}
On the other hand, for $\vec{I} = \hat{z}$, we have
$a_{0}^{1} = a_{0}^{2} = 0$.
Eliminating $a_{0}^{3}$ we find the mean field energy to be
\begin{equation}
E_{MF} = -4t \chi x + \frac{2J}{3 \sqrt{C_{2}^\prime}} x^{3/2}
\label{EmfZ}
\end{equation}
This result indicates that the boson condensate prefers to stay in the
manifold that satisfies
$|b_{1}| = |b_{2}|$, $i.e.$, $I_{z} = 0$.  As pointed out earlier, this
state is equivalent to the
$d$-wave pairing state as opposed to the state $\vec{I} =
\hat{z}$ which
corresponds to the staggered flux state with finite chemical potential.

We can follow the procedure of the last section to derive an effective
Lagrangian for the $z$
field.  The important difference is the appearance of the anisotropy
energy.  Ignoring the gradient
terms from the fermion contribution, we can write down the following
effective Lagrangian.
\begin{eqnarray}
{\cal L}_{eff} = & \frac{2}{3} & \frac{m_{b}}{D_{1}} |z^\dagger
\partial_\tau z|^{2}
                  + xz^\dagger \partial_\tau z +
\frac{x}{2m_{b}}|\partial_{i}z|^{2} \nonumber \\
               & + & \frac{x^{2}J^\prime}{2C_{1}} 4|z_{1}z_{2}|^{2}
                  + \frac{x^{2}J^\prime}{2C_{3}} \left(
                  |z_{1}|^{2} - |z_{2}|^{2} \right)^{2}
\label{Lsc}
\end{eqnarray}
The last two terms are introduced phenomenologically to model the breaking
of the O(3) symmetry
down to $x$-$y$ symmetry when $C_{1} \neq C_{3}$.  This is adequate for
$\vec{I}$
near the $x$-$y$ plane but, strictly speaking, needs further modification
near the north and south
poles, due to the singular behavior of the energy cost given by Eq. \ref{EmfZ}.

To gain some
understanding of the phases of the nonlinear $\sigma$-model, let us
consider the classical limit
where the $\tau$ dependence of $z$ are neglected.  If $(z_{1},z_{2})$ are
restricted to the
manifold of minimum energy, $i.e.$, $\vec{I}$ is in the
$x$-$y$ plane, the model
is equivalent to two $x$-$y$ models, with $K$-$T$ transition at $T_{KT}
\approx \frac{1}{4m_{b}}
\pi x$.  This temperature scale will be suppressed by fluctuations of
$\vec{I}$
out of the $x$-$y$ plane, because the energy cost per unit area is only
$x^{2}J$.  However, we need
to introduce gauge fields to Eq. \ref{Lsc} before the low lying excitations can
be fully discussed.

\section{The electron spectral function in the $\sigma$-model description}

In this section we compute the physical electron Green function $G(r,\tau)$
assuming that we are in
the disordered phase of the $\sigma$-model description.  We have
within the mean field theory
\begin{eqnarray}
G(\vec r,\tau) & = & -\frac12 \langle  T_\tau h^\dagger \psi
(\vec{r},\tau) \psi^\dagger(0,0)
                h\rangle \label{Ge} \\
          &\approx & G_{B}(\vec{r},\tau) G_F
(\vec{r}, \tau)
\label{Gmean}
\end{eqnarray}
where
\begin{eqnarray}
G_{B}(\vec{r}, \tau) & = &  \langle T_\tau 
\left( h^{\dagger} (\vec{r} , \tau)h(0,0) \right) \rangle \nonumber \\
G_{F}(\vec{r}, \tau) & = & -\langle T_\tau \left(
\tilde{\psi} (\vec{r}, \tau) \tilde{\psi}^\dagger(0,0) \right) \rangle
\end{eqnarray}

The boson Green function contains two parts.  Note that at temperature $T$
most bosons are in
states which have energies of order $T$ from the bottom of the boson band.
Thus at high energies
the boson Green function is given by the single-boson Green function
$G_{b}^{s}$ as if no other
bosons are present.  The imaginary part of this part of boson Green function
extends the whole band
width of the boson band.  At low energies (of order $T$), the boson Green
function is determined by
those nearly condensed bosons at low energies.  Thus we may assume that
bosons do condense and the
second part of the boson Green function can be approximated by 
Const.$\times e^{iQ_b \vec r}$ 
where $Q_b$ is the momentum of the bottom of the boson band.  From the above
discussion we see that the mean field electron Green function has a form
\begin{equation}
G_{e}^{(0)} = {\rm Const.}\, e^{iQ_b \vec r}
 G_{F} + G_{in}^{(0)}
\end{equation}
The second term comes from the convolution of $G_{B}^{(s)}$ and $G_{F}$ and
is the incoherent part of
the Green function.  The first term is the coherent part since its
imaginary part is given by
discrete $\delta$-functions.  (Note those discrete $\delta$-function peaks
should really have a
finite width of order $T$ if the bosons do not really condense as in the s-flux
and uRVB phases.)  It
is this coherent part that gives rise to the quasiparticle peaks observed
in photoemission experiments.  
The more exact expression of $G_{e}^{(0)}$ is given by
Eq. \ref{mGe} in Appendix C. At low temperatures the lengthy expression
can be simplified as
\begin{equation}
G_{e}^{(0)}(\omega,k) \simeq \frac{x}{2} \left[ \frac{ \left( v^{f}(k) \right)^{2}
}{ \omega-E_{-}^{f}(k)} +
                      \frac{\left( u^{f}(k) \right)^{2} }{\omega-E_{+}^{f}(k)}
\right] + G_{in}^{(0)}
\end{equation}
The incoherent part satisfies
\begin{equation}
\int_{-\infty}^{+\infty} \frac{d\omega}{\pi} \Im G_{in}^{(0)}=\frac{1}{2}
\end{equation}
which can be shown by using Eq. \ref{mGe}.

In the following we go beyond the mean field theory and discuss
several corrections to the mean field Green function.
As low energies, bosons are nearly condensed.
The boson fields $(b_1,b_2)$ (or $g_{i}(t)$) change
slowly in time direction (within a range of order $\frac{1}{T}$) and in
spatial direction (within a
range of order $a\sqrt{ \frac{t}{T} }$).
Thus locally we may think there is really a boson condensation and calculate
the (mean field) electron Green function in the boson condensed phase.
Since in the different regions, the boson fields $(b_1,b_2)$ point
to different directions, the total Green function
can be obtained by averaging the mean field Green functions for
all the directions.
We would like to point out that the fermion Green functions are different
for different directions of the boson fields, because different local
boson fields give rise to different local $a_{0}^l$
which enforces the constraint.

The above picture of calculating electron Green function naturally
comes from our $\sigma$-model treatment of the $SU(2)$ theory.
The averaging weights for different directions are determined
from the $\sigma$-model.
We now make the crude
approximation that we
are in the high temperature phase of the $\sigma$-model, where all slowly
varying configurations
$z$ are equally likely.

We have already seen in Section 3
that this procedure yields a Fermi surface which obeys
 Luttinger theorem in the Fermi
liquid phase where the bosons are condensed and in the uRVB phase
where the bosons are nearly condensed.
The fluctuations of the boson fields in the uRVB phase
will give rise to finite broadening of the
quasi-particle peaks.
We now perform the same procedure in the s-flux phase.

For each uniform configuration $g_{i} = g$, $\Im G_{F}(\omega,k)$ in general
contains four
$\delta$-function peaks as a function of $\omega$.  (Note for general $g$ we
have both translation
symmetry breaking and fermion pairing.)  After averaging over all
orientations of $g$, we get a
translation invariant electron Green function.  This averaging also gives
quasiparticle peaks an
intrinsic width and line shape.
\begin{figure}
\centerline{\epsfig{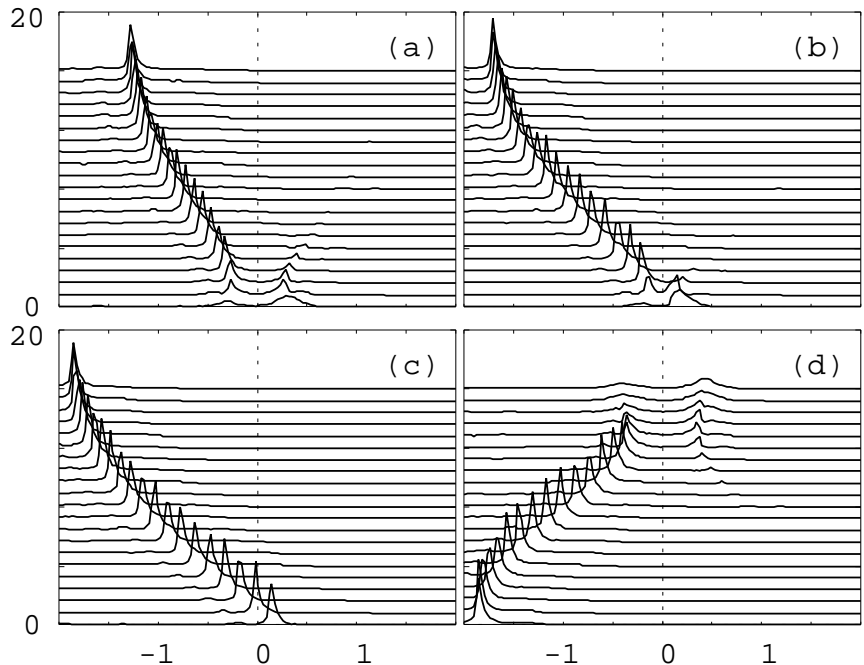}}
\caption{
The electron spectral function for, from top down,
 (a) $k=(-\pi/4, \pi/4) \to (\pi/4, 3\pi/4)$,
(b) $k=(-\pi/8, \pi/8) \to (3\pi/8, 5\pi/8)$, (c) 
$k=(0, 0) \to (\pi/2, \pi/2)$, and 
(d) $k=(0, \pi/2) \to (0,0)$.
We have chosen $J=1$.
}
\label{fig1}
\end{figure} 

Figure \ref{fig1} presents a numerical calculation of the electron spectral
function using the above
approximation. 
We have chosen $\tilde J=J/2$ and $\chi = 1$ so that
the fermion band bottom is at around $-2J$, to be consistent with experiments.  
We have set $\tilde t=t=2J$ so that the incoherent part of the spectral
function extend from $-8t=-16J$ to $0$, in order
to agree with the numerical results.   We have also  set
$\frac{\Delta}{\chi} = 0.2$ so that the gap near $(0, \pi)$ is about
$0.4 J$.
Roughly, the spectral function
is similar to that of a
$d$-wave paired state with a spin gap around $(0, \pm \pi)$ and $(\pm \pi,
0)$ of order
$\Delta_{spin} \sim 0.4 J$.  However the line shape and line width are
quite different.  If one plots
$\Im G(\omega = 0, k)$ one can see that wings toward $(0, \pi)$ and $(0, \pi)$
at two sides of the peak
(at $\left(  \frac{\pi}{2} - \delta, \frac{\pi}{2} - \delta  \right)$) are
enhanced by the
averaging.  It is because when $g = 1$, 
$\Im G(\omega = 0,k)$ has a Fermi pocket
around $\left( \pm
\frac{\pi}{2},\pm \frac{\pi}{2} \right)$.  We see that the averaging over
$g$ pushes the $d$-wave
spectrum towards a spectrum which shows a segment of Fermi surface.

In the above calculation of the spectral function, we only include some
simple fluctuations ($i.e.$
the uniform fluctuations of boson field).  One may wonder how reliable is
the above result?  In the
following we calculate $\Im G_{e}$ by including some different
fluctuations.  We find that the spin
gaps around $(0, \pm \pi)$ and $(\pm \pi,0)$ are quite robust.  However
the low energy spectral
function near $\left( \pm \frac{\pi}{2},\pm \frac{\pi}{2} \right)$
(together with the positions and
the shapes of the Fermi segments) are sensitive and are essentially
determined by the
fluctuations.  Although different fluctuations have different effects, they
in general stretch
Fermi points of the mean field theory into Fermi segments.

\begin{figure}
\centerline{\epsfig{file=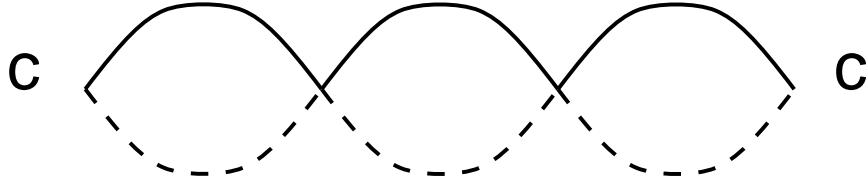, width=4.5truein}}
\caption{
A diagram for renormalized electron Green function.
The solid (dash) line is the fermion (boson) propagator.
}
\label{fig2}
\end{figure}

\begin{figure}
\centerline{\epsfig{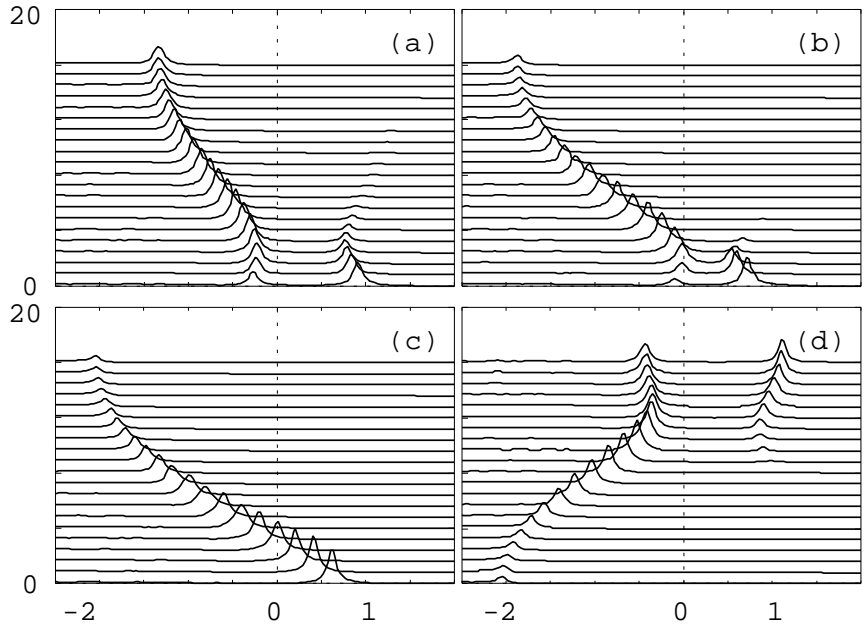}}
\caption{
The electron spectral function for, from top down, 
(a) $k=(-\pi/4, \pi/4) \to (\pi/4, 3\pi/4)$,
(b) $k=(-\pi/8, \pi/8) \to (3\pi/8, 5\pi/8)$, 
(c) $k=(0, 0) \to (\pi/2, \pi/2)$, and
(d) $k=(0, \pi) \to (0,0)$.
We have chosen $J=1$.
The paths of the four momentum scans are shown in Fig. 5
}
\label{fig3}
\end{figure}

\begin{figure}
\centerline{\epsfig{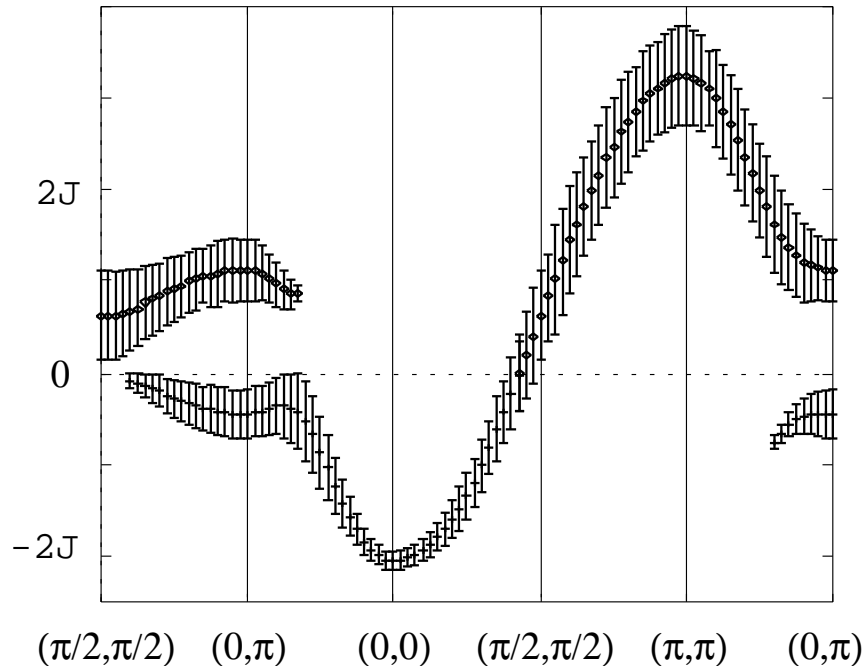}}
\caption{
The points describe the dispersion of the quasi-particle peaks for the
s-flux phase in Fig. 3.
The vertical bars are proportional
to the peak values of Im$G_U$ which are proportional to
the quasi-particle weight.
}
\label{fig3a}
\end{figure}

The dominant effect of fluctuations is to bind the bosons and the fermion
into an electron.  This
corresponds to an effective attraction between the bosons and the fermions.
One way to include
this effect is to use the diagram in Fig. \ref{fig2} to approximate the
electron Green function, which corresponds to an effective short range
interaction of form
\begin{equation}
-\frac{V}{2} (\psi^\dagger h) (h^\dagger \psi )=-V  c^\dagger c
\end{equation}
with $V < 0$.
We get
\begin{equation}
G_{e} = \frac{1}{\left( G_{e}^{(0)} \right)^{-1} + V}
\label{GV}
\end{equation}
However,
in general,  fluctuations induce more complicated interactions.
A more careful treatment can be found in Appendix C, where
we treat two different kinds of fluctuations. The first one
is the fluctuations of
$a_{0}^{\ell}$ which induces the
following interaction between the fermions and the bosons:
\begin{equation}
 \psi^{\dagger} \vec{\tau} \psi \cdot h^{\dagger}
\vec{\tau}h
\end{equation}
The second one (whose importance was pointed out by Laughlin \cite{30})
is the fluctuations of $|\chi_{ij}|$ which induces
\begin{equation}
-t (\psi^\dagger h)_j (h^\dagger \psi )_i =-2t c^\dagger_j c_i
\end{equation}
This is nothing but the original hopping term. We expect the coefficient
$t$ to be reduced due to screening, but in the following we adopt the form
\begin{equation}
V(k)=U + 2t(\cos k_x +\cos k_y)
\end{equation}
for $V$ in Eq. \ref{GV}. Here
the first and the second term comes from the first and the second kind
of fluctuations.
In Fig. \ref{fig3} and \ref{fig3a} we plot the electron spectral function for
calculated from Eq. \ref{GV}.
We have chosen $\tilde J=J/2$, $\tilde t=t=2J$,
$\chi = 1$, $\Delta/\chi = 0.4$, $x = 0.1$, and $T =
 0.1 J$. 
Here we choose $\frac{\Delta}{\chi} = 0.4$ 
so that the renormalized gap near $(0, \pi)$ is about $0.4 J$.
The value $U$ is determined from requiring the renormalized
electron Green function to satisfies the sum rule
\begin{equation}
\int_0^\infty \frac{d\omega }{2\pi}\int \frac{d^2k}{(2\pi)^2}
 \Im G_e= x
\end{equation}
Note the mean field electron Green function in Eq. \ref{mGe}
does not satisfy this sum rule:
\begin{equation}
\int_0^\infty \frac{d\omega }{2\pi}\int \frac{d^2k}{(2\pi)^2}
  \Im G_e^{(0)}= x/4
\end{equation}

\begin{figure}
\centerline{\epsfig{file=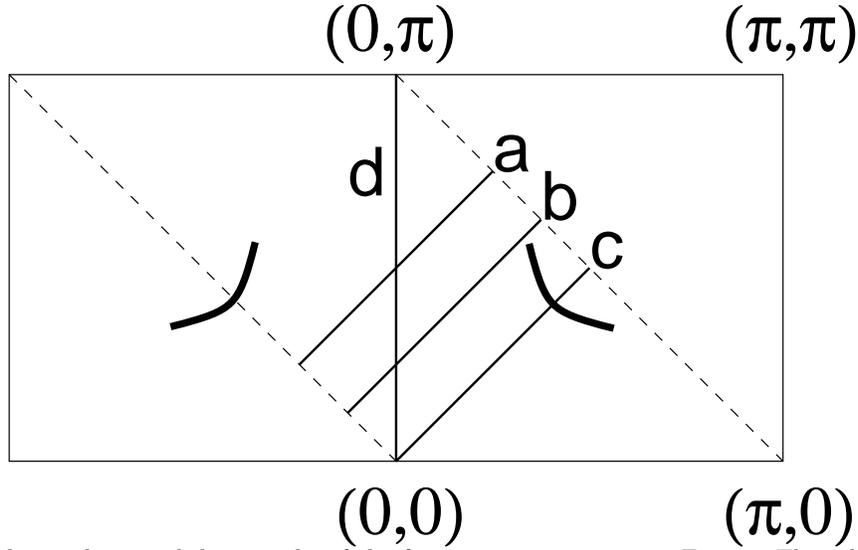, width=4.5truein}}
\caption{
The solid line a, b, c, and d are paths of the four momentum scans in Fig. 3.
The solid curves are schematic representation of
the Fermi segments where the quasiparticle peak
crosses the zero energy.
}
\label{fig7}
\end{figure}

We find that the gap near $(0, \pm
\pi)$ and $(\pm \pi,0)$ survives the inclusion of gauge and
$|\chi_{ij}|$ fluctuations.
However spectral functions
near $(\pm \frac{\pi}{2}, \pm \frac{\pi}{2})$ are modified.  The Fermi
point at $(\frac{\pi}{2},
\frac{\pi}{2})$ for the  mean field electron Green function $G_{e}^{(0)}$ is
stretched into a Fermi
segment as shown in Fig. \ref{fig7}.  We would like to point out that
the electron Green function obtained here does not show any ``shadow band''
at $\omega=0$, {\it i.e.} Im$G_{e}(0, k)$ does not have any peak
outside of the $(0,\pi)$ -- $(\pi,0)$ line as the mirror image of the peaks
that appear inside of the  $(0,\pi)$ -- $(\pi,0)$ line.

The spectral function obtained here is qualitatively similar but
quantitatively different from the one obtained in Ref. \cite{24}
through a similar calculation. The only difference is that here
we include an additional term $2t(\cos k_x +\cos k_y)$. Without
this term the quasiparticle peaks near $(0,0)$ get strongly renormalized
and become very strong. The quasiparticle energies near $(0,0)$
get pushed so high that they are nearly degenerate with the energy gap
at near $(0,\pi)$. Those features obviously disagree with experimental
observations. After including the  $2t(\cos k_x +\cos k_y)$ term
the agreement with experiments improved a lot. Due to a cancellation
between the $U$ and $2t(\cos k_x +\cos k_y)$ near $k=(0,0)$,
The quasiparticle energies and spectral weights near $(0,0)$ are quite
close to the mean field values, and the gap at $(0,\pi)$ now can be quite
different from the quasiparticle energy at $(0,0)$.

The incoherent part of the electron spectral function contains
two broad peaks, each with a width about $4t$. The incoherent part of the 
electron spectral function is roughly given by the boson density
of states. In the $SU(2)$ theory, the bosons see the staggered flux
which cause the double peak structure in the boson density
of states and in the incoherent part of the electron spectral function.
As we change $k$, the relative weight of the two broad peaks changes
due to the $k$ dependence of the coherence factors $u$ and $v$.
The mean field results of
the double-peak structure and the way in which the relative weight changes
agree quite well with the numerical calculations.\cite{34}
However the numerical 
calculations also observed certain shift of the positions of the two peaks
 as $k$ changes.
The mean field results do not have this shift. If we only include
the $U$ term the peak positions in the renormalized electron spectral function
still do not shift much. However, if we
include both the $U$ and $2t(\cos k_x +\cos k_y)$ terms
the peak positions start
to shift in the way similar to what observed in numerical calculations,
as has been pointed by Laughlin in Ref. \cite{30}.

\begin{figure}
\centerline{\epsfig{file=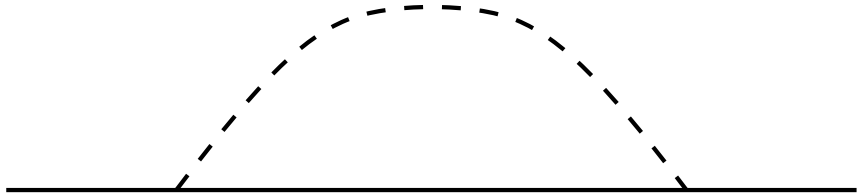, width=4.5truein}}
\caption{
A self energy diagram for the fermion Green function.
The solid (dash) line is the fermion (gauge) propagator.
}
\label{fig4}
\end{figure}

The electron spectral functions calculated above have pretty sharp peaks
even at high energies (say
for $\omega \approx -2 J$) in contrast to experimental findings that
quasiparticle peaks are much wider
at high energies.  This discrepancy can be resolved by including the self
energy of the fermion due
to the gauge fluctuation.  One can show that the self energy from the
diagram in Fig. \ref{fig4} is
proportional to $\omega$ and $k$.  Thus the lifetime is proportional to the
inverse of the quasiparticle
energy.  To include this effect, we may assume the electron Green function
to have a form
\begin{eqnarray}
G_{e}^{(0)} & \simeq & \frac{x}{2} \left[ \frac{\left(
v^{f}(k) \right)^{2}}{\omega-E_{-}^{f}(k)-i\gamma(\omega)}  
+ \frac{\left( u^{f}(k)
\right)^{2}}{\omega-E_{+}^{f}(k)-i\gamma(\omega)}
\right] + G_{in} \\
G_{e} & = & \frac{1}{\left( G_{e}^{(0)}\right)^{-1} + U }
\end{eqnarray}
If we assume the decay rate of the fermion to be $\gamma(\omega) =
|\omega| + \gamma_{0}$,  the
resulting spectral function 
is quite similar to the line shapes observed in experiments.

To summarize, we have considered three models which
treat different types of fluctuations.

\begin{enumerate}
\item \label{A.} Locally condensed boson picture.  In this picture the
quasiparticle peaks obtain
intrinsic line widths and line shapes.  Also, this picture allows us to
recover the Fermi surface
which has the Luttinger volume in the uRVB phase and the Fermi liquid
phase.

\item \label{B.} Short range attraction between the bosons and the fermions.
Those interactions are mainly due to the $a_0^\ell$
gauge and $|\chi_{ij}|$
fluctuations, which stretch the Fermi points of mean field Green
function into Fermi segments.
This attraction can also make the electron Green function to satisfy the
spectral weight sum rule of the $t$-$J$ model.

\item \label{C.} Decay of fermions $(\gamma \propto \omega)$ due to the gauge
fluctuations.  This effect
broadens the quasiparticle peaks at high energies, and makes the spectral
function look quite
similar to the ones observed in experiment.
\end{enumerate}

\section{Gauge fields}

We next investigate the low lying excitations of the effective action.  We
first consider the uRVB
state.  Starting from Eq. \ref{LuRVBb}, 
it is natural to introduce the transverse
component of the gauge
fields $\vec{a}_{i}$ by the standard replacement
\begin{equation}
\partial_{i} \rightarrow \partial_{i} + i\vec{a}_{i} \cdot
\vec{\tau} + ie \vec{A}_{i}
\label{mingaug}
\end{equation}
Recall that in the $U(1)$ case, transverse gauge field enforces the
constraint that the sum of the
fermion and boson current should vanish.  Here the three components of the
gauge field
$a_{i}^\ell$, $\ell = 1,2,3$ enforces the vanishing of the analogous
$\tau^\ell$ currents
corresponding to the $\tau^\ell$ density constraint given in Eq. \ref{cn}.  An
important difference is
that in the $SU(2)$ formulation the external electromagnetic field couples
only to the bosons,
because the physical electron density is given in terms of the boson
density by Eq. \ref{cbx}, whereas in
the $U(1)$ formulation one is free to couple the $\vec{A}$
field to the boson or
fermion, and the physical response function is the same after including the
screening by the $U(1)$
gauge field, leading to the Ioffe-Larkin combination rules.  We shall see
how these rules are
recovered or modified in the $SU(2)$ case.

In the uRVB case, it is most convenient to rotate locally to the $U(1)$
formulation as done in Eq.
\ref{LuRVB0}.  For $g_{i}$ slowly varying in space, we have
\begin{eqnarray}
U_{ij}^{(0)} & \rightarrow & g_{i}^\dagger U_{ij}^{(0)}g_{j} \nonumber \\
          & = & U_{ij}^{(0)} + i\chi_{0} (\partial_{\alpha}g_{j}^{\dagger})
 g_{j}
\label{Uijg}
\end{eqnarray}
where $i=j + \alpha$.
This is because $U_{ij}^{(0)}$ is invariant under any global rotation.  The
second term in Eq. \ref{Uijg}
gives rise to the usual transformation property of $SU(2)$ gauge fields:
\begin{equation}
\vec{a}_0^\prime \cdot \vec \tau= g^\dagger
\vec{a}_0 \cdot
          \vec{\tau} g + ( \partial_0 g^\dagger ) g,\ \ \ \ \ \
\vec{a}_i^\prime \cdot \vec \tau= g^\dagger
\vec{a}_i \cdot
          \vec{\tau} g -i ( \partial_i g^\dagger ) g
\end{equation}
after combining with Eq. \ref{a0g}.  In the rotated frame,
the fermion
$\tilde{\psi}$ obeys the $U(1)$ mean field solution with a chemical potential
which enforces the
fermion density to be $1-x$.  We can now expand to quadratic order in
$a_\mu^\prime$.  The effective
Lagrangian takes the form
\begin{eqnarray}
{\cal L}_{eff} & = & \tilde{h}^\dagger \left(
                     \partial_\tau + \vec{a}_{0}^\prime
                     \cdot \vec{\tau} + eA_{0} \right)
                     \tilde{h} + \frac{1}{2m_{b}} \left|
                     \left( \partial_{i} + i\vec{a_{i}^\prime} \cdot
                     \vec{\tau} + ieA_{i}
                      \right) \tilde{h} \right|^{2} \nonumber \\
               & - & \mu \tilde{h}^\dagger \tilde{h} + D_{1}m_{b}^{-1} \left(
                     \tilde{h}^\dagger \tilde{h} \right) ^{2} \nonumber \\
             & + & \frac{1}{2} a_\mu^{\prime \ell}(q,\omega_{n}) 
a_\nu^{\prime m} (-q,-\omega_{n})
                     \pi_{\mu \nu}^{\ell m} (q,\omega)
\label{LuRVB}
\end{eqnarray}
where $\tilde{h} = (b,0)$,
\begin{equation}
\pi_{\mu\nu}^{\ell m}(q,\omega) = \langle j_{\mu}^{\ell}(q,\omega_{n})
j_{\nu}^{m}(-q,-\omega_{n}) \rangle \;\;\;,
\end{equation}

\begin{equation}
j_{0}^{\ell} = \tilde{\psi}^{\dagger} \tau^{\ell} \tilde{\psi} \;\;\;,
\end{equation}
\begin{equation}
j_{\alpha}^{\ell} = i\left(
           \tilde{\psi}^{\dagger}\tau^{\ell}\partial_{\alpha}\tilde{\psi} -
           (\partial_{\alpha}\tilde{\psi}^{\dagger})\tau^{\ell}\tilde{\psi}
           \right)  \;\;\;.
\end{equation}
The spatial components are purely transverse
\begin{equation}
\pi_{\mu\nu}^{\ell m} = \left(  \delta_{\mu\nu} -
\frac{q_{\mu}q_{\nu}}{q^{2}}  \right)
\pi_{\perp}^{\ell m}(q,\omega)
\end{equation}
and in the uRVB state $\pi_{\perp}^{\ell m}= \delta_{\ell m}\pi_{\perp}$.  This
gives rise to three
degenerate transverse gauge modes which are massless in the uRVB state.
This is confirmed by
explicit calculation in appendix B.  In the Fermi liquid phase $b$ is Bose
condensed and these
modes are massive due to the Anderson-Higgs mechanism.

It is now clear from Eq. \ref{LuRVB} 
that only the component $a_{\mu}^{\prime(3)}$
is capable of screening
the $\vec{A}$ field.  Thus it is the component of the
gauge field parallel to the
quantization axis in the rotated frame which plays the role of the $U(1)$
gauge field. Eq. \ref{LuRVB} becomes
\begin{eqnarray}
{\cal L}_{eff} & = & b^{\dagger} \left(  \partial_{\tau} + a_{0}^{\prime 3}
+ eA_{0}     \right)
                     b + \frac{1}{2m_{b}} \left|  \left(  \partial_{i} +
a_{i}^{\prime (3)} + e A_{i}
                     \right) b \right|^{2} \nonumber \\
               & - & \mu |b|^{2} + D_{1}m_{b}^{-1} |b|^{4} \nonumber \\
                      & + & \frac{1}{2m_{b}} \left[   \left( a_{i}^{\prime (1)}
\right)^{2} + \left(
                      a_{i}^{\prime (2)} \right)^{2}
                      \right]  |b|^{2} \nonumber \\
               & + & \frac{1}{2}
 a_{\mu}^{\prime \ell} (q,\omega_{n}) a_{\nu}^{\prime \ell}
(-q,-\omega_{n})
                      \pi_{\mu\nu}^{\ell} (q,\omega)
\label{LuRVBa}
\end{eqnarray}
The coupling of the perpendicular components $\ell = 1,2$ to $b$ may be
approximated by the
expectation value of $\langle \left( a_{i}^{\prime\ell} \right)^{2} \rangle
|b|^{2}$ which simply
renormalizes the chemical potential $\mu$.  The $a_{i}^{3}$ component
can be integrated out and
gives rise to the Ioffe-Larkin combination rule
\begin{equation}
(\pi_{\mu\nu})^{-1} = (\pi_{\mu\nu}^{B})^{-1} + (\pi_{\mu\nu}^{F})^{-1}
\end{equation}
where $\pi^{F}$ is the $ \ell = 3$ component, $i.e.$ the usual
density-density response function of
the fermions.  Equation \ref{LuRVBa}
also shows that even in the $SU(2)$ formulation,
only a single component
of the gauge field couples to the boson phase and plays an important role
in suppressing the phase
coherence of the boson, just as in the $U(1)$ theory.

We next turn our attention to the s-flux phase.  The main difference is
that the $U_{ij}^{0}$
matrix is invariant only under a $\tau_{3}$ rotation $g_{i} = \exp (i\theta
\tau_{3})$ so that the
$SU(2)$ symmetry is broken down to $U(1)$.  This produces a mass
in the $\ell = 1,2$ modes and only the $\ell = 3$ gauge mode remains
massless.  This is also
checked by explicit calculation in appendix B.  Phenomenologically we are
led to the following effective Lagrangian by gauging Eq. \ref{Lsc}
 and keeping only the $a_{\mu}^{3}$ component of the gauge field.
\begin{equation}
{\cal L}_{eff} =
\frac{2m_b}{D_1} | z^\dagger D_0 z|^2
+ix z^\dagger D_0 z
- \frac{x}{2m_b}| D_i z|^2
- \frac{4 x^2 \tilde J}{2 C_1} |z_1z_2 |^2
- \frac{x^2\tilde J}{2 C_3} ( |z_1|^2- |z_2 |^2)^2
+ \frac{1}{2} a^3_\mu \pi^{\mu\nu} a^3_\nu
\label{Lsca}
\end{equation}
where $\pi^{\mu\nu}$ is the polarization tensor of the fermions
for $a^3_\mu$ gauge field, and 
$D_0=\partial_\tau +eA_0+a^3_0\tau^3$ and
$D_i=\partial_i +ieA_i+ia^3_i\tau^3$.

Equation \ref{Lsca} 
describes the low lying excitations of the underdoped regime:
the superconducting and
the spin gap phase corresponds to the ordered and thermally disordered
phases of ${\cal L}_{eff}$
respectively.  We defer a full discussion of this problem to a later
publication.  Here we give a
qualitative discussion of the superfluid density in the low temperature
phase.  We integrate out
$a_{\mu}^{3}$ in the standard way and we find the following dependence of
the transverse
electromagnetic field $\vec{A}_{i}$.
\begin{equation}
{\cal L}_{eff} = (eA_{i})^{2} \frac{x}{m_{b}} \left[
1 - \frac{(z^{\dagger}\tau^{3}z)^2}{1 + (m_b \pi_\perp /x)}
\right]
\label{LA2}
\end{equation}
where $\pi^{ij}=\pi_\perp (\delta_{ij} -\frac{q_iq_j}{q^2})$. 
The coefficient  of $A_{i}^{2}$ is the superfluid density.  On the minimum
energy manifold
$|z_{1}| = |z_{2}|$ the second term in Eq. \ref{LA2} vanishes and the superfluid
density is exactly $x$.
This is the d-wave state as we discussed earlier.  On the other hand, when
the vector
$\vec{I}$ points towards the north or south pole,
$|z_{1}z_{2}| = 0$ and we have
$z^{\dagger}\tau^{3}z = \pm 1$.  In this case the fermion spectrum is that of
the staggered flux phase
with a finite chemical potential.  The response $\pi^{\mu\nu}$ for this
orientation of $\stackrel
{\rightarrow}{I}$ is that of a metal, which vanishes in the limit $\omega,q
\rightarrow 0, \omega < q$.  In
this case the two terms in Eq. \ref{LA2} cancel and we find that $\rho_{s} = 0$,
$i.e.$ the staggered
flux phase is not a superfluid.  In general, we expect that in the
superconducting phase
fluctuations of $\vec{I}$ away from the equator will
cause a reduction of the
superfluid density due to the second term of Eq. \ref{LA2}.  For a more complete
treatment, we will need
more detailed information on $\pi^{\mu\nu}$ and its dependence on
$\vec{I}$ which
will be discussed elsewhere.

\section{Conclusion}

The main result of this paper is the derivation of the effective low energy
Lagrangian in terms of
the boson fields $z_{1}$ and $z_{2}$ and their coupling to gauge fields.
These are given in Eq. \ref{LuRVBb} together with Eq. \ref{mingaug} for the
uRVB and Fermi liquid phases,
and Eq. \ref{Lsca} for the s-flux and underdoped superconducting phases.  
In the case of the uRVB phase and the Fermi liquid
phase, we show that the
$\sigma$-model approach allows a smooth cross-over to the $U(1)$ mean field
description, recovering
all the desirable properties such as the Luttinger theorem for the Fermi
surface area and the
Ioffe-Larkin combination rules.  This is a considerable improvement over
the $SU(2)$ mean field theory.

In the staggered flux phase the $\sigma$-model description offers some new
insight into the
connection of the $SU(2)$ with the $U(1)$ theory.  The staggered flux phase is
the disordered phase of
the effective Lagrangian 
Eq. \ref{Lsca} so that we may interpret the spin gap phase as
fluctuations among d-wave
state and s-flux states and a variety of states in between.  While
the phase diagram is
quite similar to the $U(1)$ theory, the collective excitations are very
different.  In the $U(1)$
theory the gauge mode acquires a large mass gap of order $(1-x)J$.  In the
$SU(2)$ theory there are
three gauge modes, two are massive with mass of order $\Delta$, while one
remains massless.  We
believe the low lying gauge modes may help stabilize this phase.  In any
case, the massless gauge
modes will lead to large fluctuation effects which we have not truly
explored in this paper.

We also performed extensive numerical work to explore the consequence of
the $\sigma$-model
description  for photoemission experiments.  We find that within the
uncertainties of the theory
the qualitative features are not that different from the $SU(2)$ mean field
theory once the boson
fermion attraction was included.  We find an energy gap in the electron
spectrum, large near
$(0,\pi)$ and vanishing along a ``Fermi segment'' near $(\pi/2, \pi/2)$.
The precise size and
location of these segments is beyond the accuracy of the present theory, but
the
$\vec{k}$ dependence is that of a broadened
d-wave gap.  We
consider the agreement of this feature with the experiment to be strong
support of the present approach.  
{ The $SU(2)$ theory naturally describes an unusual superconducting
transition that is not associated with opening or closing of spin gap.}
We have not treated gauge fluctuations adequately in this paper
for us to describe the
energy dependence or the lineshape of the spectral function, so that at
present detailed questions
which distinguishes the energy gap as measured from the leading edge or
from the ``centroid'' of
the spectral feature remains unanswered.

We expect the $SU(2)$ mean field theory to be applicable at high temperatures
and the $\sigma$-model
description to be more accurate near the phase boundary to the
superconducting and the Fermi liquid
phases.  This is because the fermions respond to local fluctuations in the
boson fields on a length
scale of $\xi_{F} = J/T$ in the uRVB phase and $\xi_{F} = J/\Delta$ in the
s-flux phase.  On the
other hand, the boson fluctuations are on a scale $\xi_{B} = (t/T)^{1/2}$
for $T > T_{BE}^{(0)}$
and $\xi_{B} = x^{-1/2}$ for $T < T_{BE}^{(0)}$, where $T_{BE}^{(0)} = \pi
x t$ is a mean field
Bose condensation temperature.  When $\xi_{F} > \xi_{B}$, we expect the
fermion to average over the
local boson fluctuations, and the $SU(2)$ mean field theory is appropriate,
whereas the
$\sigma$-model approach requires that $\xi_{F} < \xi_{B}$.  The difficulty
is that for $T <
T_{BE}^{(0)}$ we do not have a good understanding of $\xi_{B}$ because the
coherence of the bosons
is greatly suppressed by gauge fluctuations.  In principle, we should solve
the $\sigma$ model to
obtain $\xi_{B}$ to obtain a self-consistent solution, but that is beyond
the scope of the present
paper.  This is why we explore the consequences of both methods and it is
fortunate that the
results are qualitatively similar in the s-flux phase.

One important outcome of the present work is that it is clear that the
transition to the
superconducting state is very different from the conventional BCS theory.
In BCS theory $T_{c}$ is
controlled by the closing of an energy gap in the electronic excitation
spectrum.  In the present
case, $T_{c}$ is controlled by boson fluctuations of our effective
Lagrangian.  We also note that
the effective Lagrangian is not of the conventional Ginsburg-Landau form
with a simple complex
order parameter.  The internal gauge degrees of freedom, parametrized by
$\phi$ and $\theta$ (see Eq. \ref{zphi}) play
an important role.  For example, long range phase coherence can be
destroyed by $\theta$
fluctuations.  Thus our picture of the normal phase (the disordered phase
of the $\sigma$ model) is
very different from that suggested by a number of workers, \cite{31,32,33}
 based on the idea of phase
fluctuations or a conventional BCS order parameter.  In the latter picture
normal state transport
is due to charge $2e$ collective modes, whereas we have charge $e$ metallic
carriers.  We believe
the absence of signatures of strong superconducting fluctuations in the
normal state favors our point of view.

We  would like to thank R.B. Laughlin  for sharing his unpublished results.
We also would like to thank Z.X. Shen for many helpful
discussions.
PAL is supported by NSF-MRSEC grant DMR-94-00334
and XGW is supported by NSF grant DMR-94-11574
and A.P. Sloan fellowship.

\appendix
\section{Relation between $U(1)$ and $SU(2)$ theory}

 We start with the usual $U(1)$ slave boson formalism where the operator
$c^{\dagger}_{i \sigma} $
creating an electron with spin $\sigma$
on site $i$ is represented by the spinon (fermion) operator
$f^{\dagger}_{i \sigma}$ and the holon (boson) operator
$b_i$ as
\begin{equation}
c^{\dagger}_{i \sigma} =f^{\dagger}_{i \sigma} b_i.
\end{equation}
The physical states satisfy the local constraint
\begin{equation}
\biggl(\sum_{\sigma} f^{\dagger}_{i \sigma} f_{i \sigma} + b^{\dagger}_i b_i
- 1. \biggl) | {\rm phys} > = 0
\end{equation}
Then the partition function $Z$ of the $t$-$J$ model
is represented in terms of the functional integral as
\begin{equation}
Z = \int D \psi D\psi^{\dagger} D b D b^{*} D U D a_0 e^{- \int_0^{\beta} L}
\end{equation}
with the Lagrangian $L$ being given by
\begin{eqnarray}
L &=& { \tilde J \over 2} \sum_{<ij>} {\rm Tr} \biggl[ U^{\dagger}_{ij}U_{ij} \biggr]
+ \frac{1}{2} \sum_{ij\alpha} \psi_{i\alpha}^{\dagger}
\biggl[ (\partial_\tau + \sum_{a=1}^3 a^{a}_0 \tau_a )
\delta_{ij}  + \tilde JU_{ij} \biggr] \psi_{j\alpha}
\nonumber \\
&+&   \sum_i b^{\dagger}_i ( \partial_\tau -  \mu + a^3_0 ) b_i
- t \sum_{ij} \chi_{ij} b^{\dagger}_j b_i
\nonumber \\
&=& L_F + L_B.
\label{A4}
\end{eqnarray}
where the first line is the Lagrangian $L_F$ for the fermions
while the second line is the contribution $L_B$ from the doped holes.
Here the $SU(2)$ matrix $U_{ij}$ has
the spinon pairing order parameter $\Delta_{ij}$ and the
hopping order parameter $\chi_{ij}$ as the matrix elements, i.e.,
$U_{ij} =
\left[
\begin{array}{cc}
- \chi^*_{ij} & \Delta_{ij} \\
\Delta_{ij}^* & \chi_{ij}
\end{array} \right]$.
The spinor $\psi_{i\alpha}$ is 
given by Eq. \ref{defpsi}.
We have introduced three $a_0$'s.
The 3-component $a_0^3$ is the time component of the $U(1)$ gauge field
corresponding to the constraint Eq. A2.
The 1- and 2- components
correspond to the constraint 
\begin{equation}
< {\rm phys'} | f_{i 1} f_{i 2} | {\rm phys} > =
< {\rm phys'} | f^\dagger_{i 1}f^\dagger_{i 2}| {\rm phys} > =  0
\end{equation}
which are redundant \cite{25} and are left out in the usual
$U(1)$ formulation.

Now we consider the $SU(2)$ gauge transformation which
is defined as the rotation of the
spinor $\psi_i$ in terms of a $SU(2)$ matrix $g_i$ as
\begin{eqnarray}
\psi_i &\to& {\tilde \psi}_i = g^{\dagger}_i \psi_i
\nonumber \\
U_{ij} &\to& {\tilde U}_{ij} = g^{\dagger}_i U_{ij} g_j
\nonumber \\
{ \bf a}_0 = \sum_{a= 1}^3 a_0^{a} \tau_a &\to&
{\bf {\tilde a}}_0 =
g_i^{\dagger} {\bf a}_0 g_i - (\partial_\tau g_i^{\dagger}) g_i.
\end{eqnarray}
The Lagrangian $L_F$ for the fermions remains invariant
with respect to the gauge transformation Eq. A6, while
the holon contribution $L_B$ changes.
Then away from the half-filling, $(U,{\bf a}_0)$ and
$( {\tilde U}, {\tilde {\bf a}}_0)$ are different
configuration physically.
Next we divide the functional integral over $U$ and ${\bf a}_0$
into two parts, i.e.,
the representative $(U^{(0)},{\bf a}^{(0)}_0)$
and those which are related to it by $SU(2)$
rotation $g$ as $g^\dagger U^{(0)} g$ and
${\bf a}_0 =
g_i^{\dagger} {\bf a}^{(0)}_0 g_i - (\partial_\tau g_i^{\dagger}) g_i$.
\begin{equation}
\int D U D {\bf a}_0  F(U, {\bf a}_0) =
\int D U^{(0)} D {\bf a}^{(0)}_0 \int D g F(g^\dagger U^{(0)} g,
g_i^{\dagger} {\bf a}^{(0)}_0 g_i - (\partial_\tau g_i^{\dagger}) g_i)
\end{equation}
No two members of $(U^{(0)},{\bf a}^{(0)}_0)$
are related by any $SU(2)$ rotation $g$.
We change the notation of the Grassmann variable in Eq. (A3) to
${\tilde \psi}$, and then we change the Grassmann integral
variables to $\psi = g {\tilde \psi}$,
$\psi^\dagger = {\tilde \psi}^\dagger g^\dagger$ to obtain
\begin{eqnarray}
Z = \int D {\psi} D {\psi}^{\dagger} D b D b^{*}
D U^{(0)} D {\bf a}^{(0)}_0 D g
\exp \biggl[- \int_0^{\beta} d \tau
L'(\psi, \psi^{\dagger}, b , b^{*}, U^{(0)} , {\bf a}^{(0)}_0, g ) \biggr]
\end{eqnarray}
The Lagrangian is given by
\begin{eqnarray}
L' &=& 
\frac{\tilde J}{2}\sum_{<ij>} {\rm Tr}(U_{ij}^\dagger U_{ij})
+\frac{1}{2} \sum_{ij\alpha} {\psi}^{\dagger}_{i\alpha} \biggl[
(\partial_\tau + {{\bf a}}^{(0)}_0 ) \delta_{ij}
+ \tilde J U_{ij}^{(0)} \biggr] {\psi}_{j\alpha}
\nonumber \\
&+& \sum_{ij} b_i^* \biggl[ \biggl( \partial_\tau - \mu
+ {1 \over 2} {\rm Tr}[ \tau_3( g_i^{\dagger} {{\bf a}}^{(0)}_0 g_i -
(\partial_\tau g_i^{\dagger} )g_i ) ] \biggr) \delta_{ij}
- {t \over 2} {\rm Tr} [ (1 + \tau_3) g_i^{\dagger} U_{ij}^{(0)}
g_j ] \biggr]  b_j
\end{eqnarray}
We now parameterize $g_i$ in terms of $z_i$ using Eq. \ref{gz}
and bind the $z$ with the slave boson $b$ to define
the $SU(2)$ boson $h=\pmatrix{b_1\cr b_2\cr}$ as
\begin{equation}
b_{i \alpha} = z_{i \alpha} b_i
\end{equation}
This can be represented by
\begin{equation}
h_{i} = g_i \cdot
\left[
\begin{array}{c}
b_i \\
0
\end{array} \right]
\end{equation}
Now the Lagrangian $L'$ in Eq. (A9) is written in terms of $b_1,b_2$
instead of $b$ and $g$.
First the Berry phase term is
\begin{eqnarray}
& &
\sum_{i} \biggl[ b_i^*  \partial_\tau b_i
- {1 \over 2} {\rm Tr}[ \tau_3 (\partial_\tau g_i^{\dagger} )g_i ] b^*_i b_i
\biggr]
\nonumber \\
&=& \sum_{i} \biggl[ b_i^*  \partial_\tau b_i
+ {1 \over 2} [
\sum_{\alpha} (z^*_{i \alpha} \partial_\tau z_{i \alpha}
- z_{i \alpha} \partial_\tau z^*_{i \alpha} )
 b^*_i b_i ]
\biggr]
\nonumber \\
&=& \sum_{i} \biggl[ b_i^*  \partial_\tau b_i
+ \sum_{\alpha} z^*_{i \alpha} \partial_\tau z_{i \alpha} b^*_i b_i
\biggr]
\nonumber \\
&=& \sum_{i \alpha} b_{i \alpha}^*  \partial_\tau b_{i \alpha}
\end{eqnarray}
where we have used the relation
$\sum_{\alpha} z^*_{\alpha} z _{\alpha}  = 1$ and
$\partial_\tau (\sum_{\alpha} z^*_{\alpha} z _{\alpha})
 =
 \sum_{\alpha} (
 z_{\alpha} \partial_\tau z^*_{\alpha}+  z^*_{\alpha} \partial_\tau z_{\alpha}
 ) = 0$.
Next the hopping term of the boson is written as
\begin{equation}
- {t \over 2}  {\rm Tr} \biggl[ (1 + \tau_3) g_i^{\dagger} U_{ij}^{(0)}
g_j \biggr]  b_i^* b_j = - t h^\dagger_i U^{(0)}_{ij} h_j.
\end{equation}

In summary the partition function $Z$ is written as
\begin{equation}
Z = \int D \psi D\psi^{\dagger} D h D h^{\dagger} D U^{(0)} D {\bf a}^{(0)}_0
e^{- \int_0^{\beta} {\tilde L}}
\end{equation}
with the Lagrangian ${\tilde L}$ being given by
\begin{eqnarray}
{\tilde L} &=&
{\tilde J \over 2}\sum_{<ij>} {\rm Tr} \biggl[ U^{(0) \dagger}_{ij}U^{(0)}_{ij}
\biggr]
+ \frac{1}{2}\sum_{ij\alpha} \psi_{i\alpha}^{\dagger}
\biggl[ (\partial_\tau + {\bf a}^{(0)}_0 )
\delta_{ij}  + \tilde JU^{(0)}_{ij} \biggr] \psi_{j\alpha}
\nonumber \\
&+&   \sum_i h^{\dagger}_i ( \partial_\tau - \mu  + {\bf a}^{(0)}_0
) h_i
- t \sum_{ij} h^{\dagger}_i U^{(0)}_{ij} h_j.
\label{A15}
\end{eqnarray}
Now the Lagrangian is invariant with respect to the $SU(2)$ gauge transformation
given in Eq. \ref{Gtran}.
Then the constraint that no two configuration $(U^{(0)},{\bf a}_0^{(0)})$
are related by $g$ can be relaxed because it gives only the constant gauge
volume. Then we can drop $(0)$ from $U$ and ${\bf a}_0$.
This has the form of the $SU(2)$ gauge theory proposed by Wen and Lee.
However, we note that in the latter theory, the last term
in Eq. \ref{A15} is replaced by 
$\tilde t \sum_{ij} h^{\dagger}_i U^{(0)}_{ij} h_j$ where
$\tilde t=t/2$. A possible source of this difficulty is that in
Eq. \ref{A4} we impose three constraints using three Lagrangian
multipliers $a_0^a$, whereas in the standard $U(1)$ formulation,
only a single  Lagrangian
multiplier $a_0$ is used. We cannot justify this procedure because
the three constraints involve noncommuting operators.
Another possible source of discrepancy is that in going from integration
over $b$ and $g$ to integration
over $h$, a Jacobian may be necessary.

\section{Microscopic derivation of Gauge Fields}

In this Appendix, we describe the microscopic derivation of the gauge fields
in each of the mean field states. 
We begin by giving several arguments for when the gauge field is
expected to be massless. We then show by explicit calculation
that for the uRVB phase there are three
massless transverse gauge fields. Finally we present a
calculation of the propagator of the massless gauge field in the
s-flux phase after integrating out the fermions.
Because we are interested in the low
energy dynamics, we consider only the massless gauge fields. The first
task is to identify the massless gauge fields.
For this purpose let us consider the following gauge-invariant term which
appear in the free energy \cite{35}.

\begin{equation}
 F = {\rm Tr}( P_{ij .....k} U_{i i'} P^\dagger_{i' j' ..... k'} U_{i' i})
\end{equation}
where
\begin{equation}
 P_{ij .....k} = U_{i j} U_{j l} \cdot \cdot \cdot U_{mk} U_{k i}
\end{equation}
is the product  of $U$'s along a closed loop $i \to j \to \cdot
\cdot \cdot \to k \to i$. When we write $U_{ij}$ as
\begin{equation}
  U_{i j} =
U^{(0)}_{i j}
e^{  i {\bf a}_{ij}}
=U^{(0)}_{i j}
e^{  i a^a_{ij} \tau_a}
\end{equation}
with $U_{ij}^{(0)}$ being the mean field configuration, we obtain the
following contribution to the free energy of $a^a_{ij}$.
\begin{eqnarray}
 \delta F &=& {\rm Tr}( P^{(0)}_{ij .....k} U^{(0)}_{i i'}
e^{  i a^a_{ii'} \tau_a}
P^{(0) \dagger}_{i' j' ..... k'}
e^{ - i a^a_{ii'} \tau_a}
 U^{(0)}_{i' i})
\nonumber \\
&=& {\rm Tr}(
 U^{(0)}_{i' i}
 P^{(0)}_{ij .....k} U^{(0)}_{i i'}
e^{  i a^a_{ii'} \tau_a}
P^{(0) \dagger}_{i' j' ..... k'}
e^{ - i a^a_{ii'} \tau_a})
\end{eqnarray}
  Then if $P^{(0)}$ does not commute with $\tau_a$, Eq. (B4) gives the mass
to the gauge field $a^a$. For example, if
$P^{(0)}= c_0 1 + c_3 \tau_3$ with $c$'s being constants,
\begin{eqnarray}
  e^{i {\bf a} } P^{(0)} e^{-i {\bf a} }
&=&   P^{(0)}
+ i [ {\bf a}, P^{(0)} ]
+{ 1 \over 2}  i^2 [ {\bf a},  [ {\bf a}, P^{(0)} ]] + ......
\nonumber \\
&=& c_0 1 + c_3 \tau_3 + c_3 ( a^1 \tau_2 - a^2 \tau_1 )
- { {c_3} \over 2} ( (a^1)^2 + (a^2)^2 ) \tau_3 + .... ,
\end{eqnarray}
and $a^1$ and $a^2$ becomes massive.
This is nothing but the Higgs mechanism where $P$ is the Higgs field
which are site variable
belonging to the adjoint (vector) representation of $SU(2)$. The condensation of
$P$ breaks the symmetry from $SU(2)$ to $U(1)$ and only one gauge field,
i.e., $a^3$ in the above example, remains massless.
On the other hand, if $P^{(0)} = c_0 1$ for every elementary plaquette,
$P^{(0)}$ for arbitrary closed loop is const $\times$ 1
independent of the  gauge choice. In this case we can choose a gauge where
$U_{ij}^{(0)} \propto  1$, and all the gauge fields $a^1,a^2,a^3$ remain
massless.
Now we apply the general consideration above to the each mean
field state. We chose the gauge where the link variable $U_{ij}^{(0)}$
for each state is given by
\begin{equation}
 U^{(0)}_{i i+x}
= U^{(0)}_{i i+y } = i \chi_0 1
\end{equation}
for uRVB state, while
\begin{eqnarray}
U^{(0)}_{i i+x}
&=& - \chi \tau_3 - i (-1)^{i_x + i_y} \Delta
\nonumber \\
U^{(0)}_{i i+y}
&=& - \chi \tau_3 +  i (-1)^{i_x + i_y} \Delta
\end{eqnarray}
for  s-Flux  and $\pi$-Flux
 states.
Then the product of $U$'s along an elementary plaquette
$P^{(0)}_{\rm pl}$ is obtained as
\begin{equation}
P^{(0)}_{\rm pl} = \chi_0^4 1
\end{equation}
for  uRVB  state, while
\begin{equation}
P^{(0)}_{\rm pl} = [ (\chi^2 - \Delta^2)^2 - 4 \chi^2 \Delta^2 ]1
 \pm 4 i \chi \Delta ( \chi^2 - \Delta^2) \tau_3
\end{equation}
for  s-Flux  and $ \pi$-Flux states.
Then it can be easily seen that in uRVB and $\pi$-Flux ( $\chi = \Delta$)
states, all the gauge fields remain massless while only $a^3$ remains
massless in s-Flux state. For the $\pi$-Flux state, we can chose the gauge
where
\begin{eqnarray}
 U^{(0)}_{i i+x}
&=& i (-1)^{i_y} \chi 1
\nonumber \\
U^{(0)}_{i i+y}
&=& i \chi 1.
\end{eqnarray}

Now we explicitly derive the effective action for the gauge fields up to the
quadratic orders.
We start from the Lagrangian in Eq. 6. We divide the link variable
$U_{ij}$ and ${\bf a}_0$ into the mean field value and the fluctuation
around it.

\begin{eqnarray}
 U_{ij}
&=&U^{(0)}_{ij} + \delta U_{ij}
\nonumber \\
{\bf a}_0
&=& {\bf a}_0^{(0)} + \delta {\bf a}_0
\end{eqnarray}
Integrating out the fermions and bosons,
we obtain the effective action for $\delta U_{ij}$ and
$\delta {\bf a}_0$.
\begin{eqnarray}
 S_{\rm eff}
&=&
 \tilde J \sum_{<ij>} {\rm Tr} [
(U^{(0) \dagger }_{ij}  + \delta U_{ij}^\dagger )
(U^{(0)}_{ij}  + \delta U_{ij}  ) ]
\nonumber \\
&-& {\rm Tr}_F ln ( - G_{F 0}^{-1} +  \delta {\bf a}_0 + \tilde J \delta U_{ij} )
+ {\rm Tr}_B ln ( - G_{B 0}^{-1} +  \delta {\bf a}_0 + t \delta U_{ij} )
\end{eqnarray}
where $\tilde J = 3J/8$ and ${\rm Tr}_F$ and ${\rm Tr}_B$ are the
fermionic and bosonic traces.
The Green's functions $G_{F0}$ and $G_{B0}$ in the mean field state  are
given by
\begin{eqnarray}
 G_{F0}^{-1}
&=&  i \omega_n -  {\bf a}^{(0)}_0 - \tilde J U^{(0)}_{ij}
\nonumber \\
 G_{B0}^{-1}
&=&  i \omega_l - \mu_B -  {\bf a}^{(0)}_0 - t U^{(0)}_{ij}.
\end{eqnarray}

Now we can expand Eq. (B12) as
\begin{eqnarray}
S_{\rm eff}
&=& S_0 + S_1 + S_2+ .....
\nonumber \\
&=&
\tilde J \sum_{<ij>} {\rm Tr} [
U^{(0) \dagger}_{ij} U^{(0)}_{ij} ]
\nonumber \\
&+&
\tilde J \sum_{<ij>} {\rm Tr} [
U^{(0) \dagger}_{ij} \delta U_{ij}
+\delta U^{\dagger}_{ij} U^{(0)}_{ij}
]
\nonumber \\
&+& {\rm Tr} [ G_{F0} (  \delta {\bf a}_0 + \tilde J \delta U_{ij}) ]
- {\rm Tr} [ G_{B0} (  \delta {\bf a}_0 + t \delta U_{ij}) ]
\nonumber \\
&+&
\tilde J \sum_{<ij>} {\rm Tr} [
\delta U^{ \dagger}_{ij} \delta U_{ij} ]
\nonumber \\
&+&{ 1 \over 2}  {\rm Tr}_F [
G_{F0} (  \delta {\bf a}_0 + \tilde J \delta U_{ij})
G_{F0} (  \delta {\bf a}_0 + \tilde J \delta U_{i'j'})
 ]
\nonumber \\
&-& { 1 \over 2}  {\rm Tr}_B [
G_{B0} (  \delta {\bf a}_0 + t \delta U_{ij})
G_{B0} (  \delta {\bf a}_0 + t  \delta U_{i'j'})
 ] + .....  .
\end{eqnarray}
The mean field equation is obtained from the condition
that the first order terms in $\delta U$ vanish.
( Here we chose the form Eq. (B7) for the s- and $\pi$-Flux
states, but the obtained mean field  equations are valid also for
uRVB state by putting $\Delta=0$ and $\chi=\chi_0$.)
\begin{eqnarray}
\tilde J\chi &=&\int_{-\pi}^{\pi} {{ d^2 k} \over {(2 \pi)^2 }}
\biggl[
{ { \tilde J \chi \gamma_k^2} \over  {4 E_k} } [ 1 - 2 f(2 \tilde J E_k)]
+{ { t \chi \gamma_k^2} \over {4 E_k} } [ n(- \mu_B - 2t E_k) -
n(- \mu_B + 2 t E_k)]
\biggr]
\nonumber \\
\tilde J \Delta &=&\int_{-\pi}^{\pi} {{ d^2 k} \over (2 \pi)^2}
\biggl[
{{ \tilde J \Delta  \eta_k^2} \over { 4 E_k} } [ 1 - 2 f(2 \tilde J E_k)]
+{{ t \Delta \eta_k^2} \over { 4 E_k} } [ n(- \mu_B - 2t E_k) - n(- \mu_B + 2 t
E_k)]
\biggr]
\end{eqnarray}
where
\begin{eqnarray}
\gamma_k = \cos k_x + \cos k_y
\nonumber \\
\eta_k = \cos k_x - \cos k_y
\nonumber \\
E_k = \sqrt{ (\chi \gamma_k)^2 + ( \delta \eta_k)^2}
\end{eqnarray}
and
$f(x)$ and $n(x)$ are the Fermi and Bose distribution functions, respectively.
The condition that the first order terms in $\delta {\bf a}_0$ vanish gives
Eq. \ref{cn}, which is satisfied by the mean field solutions.

Now we study the second order terms $S_2$. From the considerations given above
we consider only the gauge fields which commute with $P^{(0)}$.
First consider the uRVB and $\pi$-Flux states, where
$U^{(0)}_{ij} \propto  1$ and all the gauge fields remain massless.
When we make a gauge transformation
\begin{equation}
U^{(0)}_{ij}  \to
U_{ij} = g_i U^{(0)}_{ij}  g_j^\dagger
\end{equation}
the action does not change.
Let us take
 $g_i =  e^{ i \theta_i  \tau_a}$.
If $g_i$ commutes with $U_{ij}^{(0)}$, we have
\begin{equation}
U_{ij} = U_{ij}^{(0)} e^{ i (\theta_i- \theta_j)  \tau_a}
\end{equation}
If we consider this as an expansion $\delta U_{ij}$ about $U_{ij}^{(0)}$
which corresponds to a 
pure gauge configuration,
we can see that the coefficient of the second order term in 
$a^a_{ij} = \theta_i - \theta_j$
vanish for any $a=1,2,3$ for the uRVB and $\pi$-Flux phases,
and only for $a=3$ for the s-Flux phase.

Generally the second order contribution $S_2$ can be written as
\begin{equation}
S_2 = { 1 \over 2} \sum_{a,b} \sum_{ \mu, \nu}
 \sum_{ q, i\omega_n}
\Pi^{ab}_{\mu \nu}(q, i\omega_n)
a^a_{\mu}(q, i \omega_n)
a^b_{\nu}(-q, -i \omega_n)
\end{equation}
and the above consideration guarantees the masslessness of $a^a$, and
leads to the condition
\begin{equation}
\Pi^{a b}_{\mu \nu}(q=0, i\omega_n=0) =0.
\end{equation}
Here $ a,b = 1,2,3$ for the uRVB and $\pi$-Flux states while
$a=b=3$ for the s-Flux state.

To make things more clear, we describe here the explicit calculation for
the uRVB state.
Taking the gauge choice of Eq. (B6),  the mean field equation is obtained as
\begin{equation}
\tilde J\chi_0 ={1 \over 2} \int_{-\pi}^{\pi} {{ d^2 k} \over (2 \pi)^2}
\biggl[
\tilde J  {\tilde \gamma}_k f(-2 \tilde J\chi_0 {\tilde \gamma}_k )]
+ t {\tilde \gamma}_k n(- \mu_B - 2t \chi_0 {\tilde \gamma}_k) \biggr]
\end{equation}
where
${\tilde \gamma}_k = \sin k_x + \sin k_y$.
At first glance this appears different from Eq. (B15),
but it  can be shown  by using partial integration that
Eq. (B15) is reduced to Eq. (B21) by putting $\Delta=0$ and $\chi = \chi_0$.
This can be also written as
\begin{equation}
\tilde J\chi_0 = -\int_{-\pi}^{\pi} {{ d^2 k} \over (2 \pi)^2}
\biggl[
\tilde J^2 \chi_0  {\tilde \gamma}_k^2  f'(-2 \tilde J\chi_0 {\tilde \gamma}_k)
+ t^2  \chi_0 {\tilde \gamma}_k^2  n'(- \mu_B - 2t \chi_0
{\tilde \gamma}_k) \biggr],
\end{equation}
where $f'(x) = \partial f(x)/ \partial x$
($n'(x) = \partial n(x)/ \partial x$ ).

Now we consider the second order contribution $S_2$.
The gauge fields are related to $\delta U_{ij}$ as
\begin{equation}
\delta U_{i i+ \mu} = \delta U_{i i + \mu }^\dagger =
- \chi_0 a_{i i + \mu}^{a} \tau_a
\end{equation}
Then the coupling $S_{\rm int}$ between the
fermions (bosons) with the gauge field is
written as
\begin{equation}
S_{\rm int} = - { 1 \over {\sqrt{\beta N}} }
\sum_{ a, \mu \nu}\sum_{ k q}
\cos k_\mu  \biggl[
2 \tilde J\chi_0
a^a_\mu (q) \psi_{k + q/2} \tau_a \psi_{k-q/2}
+ 2 t \chi_0
a^a_\mu (q) h_{k + q/2} \tau_a h_{k-q/2}
\biggr]
\end{equation}
where $\psi,\psi^{\dagger}$ are spinors.
Now $S_2$ is explicitly given by
\begin{equation}
S_2 = \sum_{a, \mu \nu} \sum_q [ 2 \tilde J \chi_0^2 \delta_{\mu \nu}
- \Pi^{F a b}_{\mu \nu} (q)
- \Pi^{B a b}_{\mu \nu} (q)
]
 a^a_\mu (q) a^b_\nu (-q)
\end{equation}
where
\begin{equation}
\Pi^{F a b}_{\mu \nu} (q)
= 4 \delta_{ab}  \tilde J^2 \chi_0^2
\int_{-\pi}^{\pi} {{ d^2 k} \over (2 \pi)^2}
\cos k_\mu \cos k_\nu
{ { f(\xi_{k+q/2}) - f(\xi_{k-q/2})} \over
{ i \omega_m - \xi_{k+q/2} + \xi_{k-q/2} } }
\end{equation}
where $\xi_k = - 2 \tilde J \chi_0 {\tilde \gamma}_k$.
Similar expression is obtained for
$\Pi^{B a b}_{\mu \nu} (q)$.
It can be easily seen that in the limit $q \to 0$,
$\Pi^{F a b}_{\mu \nu} (q) + \Pi^{B a b}_{\mu \nu} (q)
\to 2 \tilde J \chi_0^2 \delta_{\mu \nu}$ by using the
mean field equation Eq.  (B22).
The similar cancellation is obtained for $a^1, a^2, a^3$
in the $\pi$-Flux state, and for $a^3$ in the s-Flux state.

Finally we present a calculation of the gauge field propagator
when the fermions are integrated out, {\it ie} we compute
$\Pi^{Fa b}_{\mu \nu}(q, i\omega_n) =
\Pi^{Fa b}_{\mu \nu}(q, i\omega_n)-
\Pi^{Fa b}_{\mu \nu}(q=0, i\omega_n=0)
$
in terms of the continuum approximation
in the limit of small $v_Fq, \omega_n$, and $T$ compared with
$J$.
For the uRVB state, this calculation is exactly the same as in the $U(1)$
case described in ref.10.
For the s-Flux state, we consider the following effective
Lagrangian for the fermions  in the continuum approximation:
\begin{eqnarray}
L &=& \int d^2 r \psi^\dagger_1 [ D_\tau
- 2 i \tilde J \chi
( D_x + D_y  )
\sigma_3 \tau_3
+ i 2 \tilde J \Delta
( D_x  - D_y  )
\sigma_2 ] \psi_1
\nonumber \\
 &+& \int d^2 r \psi^\dagger_2 [ D_\tau
- 2 i \tilde J \chi
( D_x  - D_y  )
\sigma_3 \tau_3
+ i 2 \tilde J \Delta
( D_x + D_y  )
\sigma_2 ] \psi_2
\end{eqnarray}
where $D_{\mu} = \partial_\mu + i\tilde a^3_{\mu}$.
Since the problem has relativistic symmetry, it is convenient
to introduce $\tilde a_{\mu}=(-i a_0, a_x,a_y)$.
The spinor $\psi_1$ ($\psi_2$) describes the fermions near $\pm (\pi/2, \pi/2)$
($ \pm (\pi/2, -\pi/2)$) in the $k$-space.
Because of the double periodicity in the gauge choice of Eq. (B7),
$k$ and $k+(\pi,\pi)$ are coupled and $\sigma$'s are the
Pauli matrices describing this $2 \times 2$ space in addition to the
original isospin space spanned by $\tau$ matrices.
When  we integrated over the fermions in Eq. (B27),
the following integral $g_{\mu \nu}({\bf q})$ appears.
\begin{equation}
g_{\mu \nu}({\bf q}) = \int {d^3 k \over {(2 \pi )^3}}
{ { k_{\mu} ( {\bf k}+ {\bf q})_\nu} \over { {\bf k}^2 ( {\bf k}+{\bf q})^2} }
\end{equation}
where ${\bf k}=(k_0,k_1,k_2)
= ( \omega, \vec k)$ is the vector in (2+1)-dimensions.

By using the  Feynman's trick, i.e,
\begin{equation}
{ 1 \over {ab} } = \int_0^1
{ {d z} \over {[ a z + b (1-z)]^2} },
\end{equation}
the integral is transformed as
\begin{eqnarray}
g_{\mu \nu}({\bf q}) &=&
= \int_0^1  d z \int 
{d^3 k \over {(2 \pi)^3}}
{ { k_{\mu} ( {\bf k}+ {\bf q})_\nu} \over
{ [({\bf k}+z{\bf q})^2 + z(1-z) {\bf q}^2]^2   } }
\nonumber \\
&=&  \int_0^1  d z \int 
{d^3 k \over {(2 \pi)^3}}
{ { k_{\mu}^2 \delta_{\mu \nu} + z(1-z) q_\mu q_\nu  } \over
{ [{\bf k}^2 + z(1-z) q^2 ]^2 }}
\end{eqnarray}
Now
$g_{\mu \mu}(q)$ is diverging if the ultraviolet cut-off $\Lambda$
for $k$-integration is infinity.
This is cured if one consider
$g_{\mu \nu}(q) - g_{\mu \nu}(0)$, which is converging when
$\Lambda \to \infty$.
Using 
$\int {d^3 k \over {(2 \pi)^3 }}
{1 \over {{\bf k}^2 ({\bf k} + {\bf q})^2}}
= {1 \over {8q}}$
where $q=\sqrt{ \vec q^2 +\omega^2}$,
we obtain
\begin{equation}
g_{\mu \nu}({\bf q}) - g_{\mu \nu}(0)
= { { q_\mu q_\nu - \delta_{\mu \nu} {\bf q}^2} \over {8q}}
\int_0^1 d z \sqrt{ z(1-z)}
= { { q_\mu q_\nu - \delta_{\mu \nu} {\bf q}^2} \over {8q}}
\end{equation}
Using Eqs. (B31) and (B32) 
we obtain the following effective action of the gauge field
at zero temperature.
\begin{equation}
S_{\rm eff} = \sum_{\bf q} {1 \over q}
\biggl[\tilde{J}^2 \chi \triangle f_{xy}({\bf q})f_{xy}(-{\bf q})
+{\chi^2 + \triangle^2 \over {16 \chi \triangle}}
(f_{0x}({\bf q}) f_{0x}(-{\bf q})
+ f_{0y}({\bf q}) f_{0y}(-{\bf q}))\biggl]
\end{equation}
where $f_{\mu\nu}=\partial_\mu \tilde a_\nu -\partial_\nu \tilde a_\mu$.
The coefficient of $ a_\mu a_\nu$ of this expression gives the inverse
of the gauge propagator which is correct for small $q$ and $\omega$ in the lattice model.

\section{Electron spectral function}

The more exact expression of mean field electron Green function
$G_{e}^{(0)}$ is given by
\begin{eqnarray}
G_{e}^{(0)} & \equiv & \langle  T \left(  c_{\uparrow}c_{\uparrow}^{\dagger}
\right) \rangle =
                        \frac{1}{N} \frac{1}{2} \sum_{q}
                        \left[  u^{b}(q-k)u^{f}(q) + v^{b}(q-k)v^{f}(q)
\right]^{2} \nonumber \\
            &        &  \times \sum_{\alpha = +,-} \left[  n_{b}\left(
E_{\alpha}^{b}(q-k) \right) +
                        n_{F}\left(E_{\alpha}^{f}(q)\right) \right]
                        \frac{1}{\omega-\left[
E_{\alpha}^{f}(q)-E_{\alpha}^{b}(q-k) \right] - i\delta}
                     \nonumber    \\
            &        &   + \frac{1}{N}\frac{1}{2}\sum_{q} \left[
                         u^{b}(q-k)v^{f}(q)-v^{b}(q-k)u^{f}(q) \right]^{2}
 \nonumber \\
            &        &  \times \sum_{\alpha = +,-} \left[  n_{b}\left(
E_{\alpha}^{b}(q-k) \right) +
                         n_{F}\left( E_{-\alpha}^{f}(q) \right)  \right]
                        \frac{1}{\omega-\left[
E_{\alpha}^{f}(q)-E_{-\alpha}^{b}(q-k)  \right] - i\delta}
\label{mGe}
\end{eqnarray} 
Here $N$ is the number of sites and
\begin{eqnarray*}
E_{\pm}^{f}(k)   & = &  \pm \sqrt{\left( \epsilon_{k}^{f}  \right)^{2} +
\left( \eta_{k}^{f}
                        \right)^{2}}  \\
\epsilon_{k}^{f} & = &  2\tilde J \chi (\cos k_{x} + \cos k_{y}) \\
\eta_{k}^{f}     & = &  2\tilde J \Delta (\cos k_{x} - \cos k_{y}) \\
E_{\pm}^{b}(k)   & = &  \pm \sqrt{\left( \epsilon_{k}^{b}  \right)^{2} +
\left( \eta_{k}^{b}
                        \right)^{2}} -\mu_B \\
\epsilon_{k}^{b} & = &  2\tilde t \chi (\cos k_{x} + \cos k_{y}) \\
\eta_{k}^{f}     & = &  2\tilde t \Delta  (\cos k_{x} - \cos k_{y}) \\
u^{f,b}(k)       & = &  \frac{1}{\sqrt{2}}
\sqrt{1 + \frac{\epsilon_{k}^{f,b}}{\sqrt{(\epsilon_{k}^{f,b})^{2} 
+ (\eta_{k}^{f,b})^2}}} \\
v^{f,b}(k)       & = &   \frac{1}{\sqrt{2}}
\frac{\eta_{k}^{f,b}}{|\eta_{k}^{f,b} |} \sqrt{1 -
\frac{\epsilon_{k}^{f,b}}{\sqrt{(\epsilon_{k}^{f,b})^{2} + (\eta _{k}^{f,b})^2}}}
\end{eqnarray*}
$n_{b,f}(E)$ are boson or fermion occupation numbers at energy $E$. 
The incoherent part comes from the terms containing $n_f(E_{\pm}^{f})$.
One can show that $\Im G_{in}^{(0)}=0$ for $\omega >0$ and
\begin{equation}
\int \frac{d\omega}{2\pi} \Im G_{in}^{(0)}=\frac{1}{2}
\end{equation}
The coherent parts come from
the terms containing $n_{b}(E_{-}^{b})$.  Note $n_{b}\left( E_{-}^{b}(k)
\right)$ is almost zero
except near $(0,0)$ and $(\pi ,\pi)$.  Approximating those peaks by
$\delta$-functions in
$k$-space, we get
\begin{equation}
G_{e}^{(0)}(k) \simeq \frac{x}{2} \left[ \frac{ \left( v^{f}(k) \right)^{2}
}{ \omega-E_{-}^{f}(k)} +
                      \frac{\left( u^{f}(k) \right)^{2} }{\omega-E_{+}^{f}(k)}
\right] + G_{in}
\end{equation}

Next we will consider effects of fluctuations.
We will consider two different types of fluctuations.
The first one
is the fluctuations of
$a_{0}^{\ell}$ whose effect is modeled by the
following effective short range
interaction between the fermions and the bosons:
\begin{equation}
 \psi^{\dagger} \vec{\tau} \psi h^{\dagger} \vec{\tau}h
\label{BFint}
\end{equation}
The second one 
is the fluctuations of $|\chi_{ij}|$ which induces
\begin{equation}
-t (h^\dagger \psi )_i(\psi^\dagger h)_j =-2t c_i c^\dagger_j
\end{equation} 

In the s-flux phase the electron operator $c^{\dagger} = \frac{1}{\sqrt{2}}
\psi^{\dagger}h$ mixes with
an operator $\tilde{c}^{\dagger} = \frac{1}{\sqrt{2}}
\psi^{\dagger}\tau^{3}h$.  We find
$\langle T \tilde{c}_{k+Q}^{\dagger} c_{k}^{\dagger} \rangle \equiv
iG_{Q}^{(0)}$ with $Q = (\pi,
\pi)$ is nonzero and given by
\begin{eqnarray}
G_{Q}^{(0)}(k) & = & \frac{1}{N} \frac{1}{2} \sum_{q} \left[
u^{f}(q)u^{b}(q-k) + v^{f}(q)v^{b}(q-k) \right] 
                 \times \left[ v^{f}(q)u^{b}(q-k) - u^{f}(q)v^{b}(q-k)
\right]  \nonumber \\
               &   &   \sum_{\alpha,\beta=+,-} \alpha \beta \left[  n_{b}
\left( E_{\beta}^{b}(q-k)
                        \right) + n_{F}\left( E_{\alpha}^{f}(q) \right) \right]
                         \frac{1}{\omega- \left[ E_{\alpha}^{f}(q) -
                          E_{\beta}^{b}(q-k) \right] -i\delta} \\
& &\simeq \frac{x}{2} v^{f}(k)u^{f}(k) \left( - \frac{1}{\omega-E_{+}^{f}(k)} +
   \frac{1}{\omega-E_{-}^{f}(k)}  \right) + \tilde{G}_{in}
\end{eqnarray} 
Since the interaction couples to both $c$ and $\tilde{c}$, we have to
invert a two-by-two matrix
to calculate the electron Green function.  Noticing that $\langle
c_{k}c_{k} \rangle =
\langle \tilde{c}_{k}^{\dagger} \tilde{c}_{k}^{\dagger}\rangle$ and introducing
\begin{equation}
\mbox{\boldmath $G$} = \left(
\begin{array}{ll}
G_{e}^{(0)}(k) & -  iG_{Q}^{(0)}(k) \\
G_{Q}^{(0)}(k) &   G_{e}^{(0)}(k+Q)
\end{array}
\right)
\end{equation}
we find that the electron Green function is the (1,1) component of the
two-by-two matrix
\begin{eqnarray}
G_e= \left[ \mbox{\boldmath $G$}^{-1} +
\left(
\begin{array}{ll}
U_{1} & 0 \\
0     & U_{2}
\end{array}  \right) \right]^{-1}_{11}
\label{Gex}
\end{eqnarray}
Note that when $U_{2} = 0$ the above equation reduces to Eq. \ref{GV}.
$U_{1,2}$ are obtained by
rewriting the interaction Eq. \ref{BFint} in the $c$ and $\tilde{c}$ basis:
\begin{eqnarray}
U \psi^{\dagger} \vec{\tau} \psi h^{\dagger}
\vec{\tau}h
& = & 3U c^{\dagger}c - U \frac{1}{2} \psi^{\dagger}
\vec{\tau}h
         h^{\dagger}\vec{\tau} \psi \\
& = & 3U c^{\dagger}c - U \tilde{c}^{\dagger}\tilde{c} + ...
 \end{eqnarray}  
The ``...'' term has a form $\psi^{\dagger}\tau^{1,2}h 
h^{\dagger}\tau^{1,2}\psi$ and does
not contribute to the electron Green function. Thus $U_{1} = 3U$ and
$U_{2} = -U$ for the
interaction in Eq. \ref{BFint}.

At low energies the interactions are dressed by fermion bubbles.  Because
$a_{\mu}^{3}$ gauge field
is massless in the s-flux phase, its fluctuations mediate a long range
interaction.  Thus the
interaction $\psi^{\dagger}\tau^{3}\psi h^{\dagger}\tau^{3}h$ is
enhanced at low energies.  To
study this effect let us consider an extreme case which has the following
interaction
\begin{eqnarray}
U\psi^{\dagger}\tau^{3}\psi h^{\dagger}\tau^{3}h \\
= Uc^{\dagger}c + U \tilde{c}^{\dagger}\tilde{c} + ...
\label{BFint3}
\end{eqnarray}
The ``...'' term does not contribute to the electron Green function.  Thus the
electron Green function 
in this case is given by Eq. \ref{Gex} with $U_{1} = U_{2} = U$.

>From the above discussion, it is also easy to see that the interaction
induced by the $|\chi_{ij}|$ fluctuations only modify $U_1$:
\begin{eqnarray}
G_e= \left[ \mbox{\boldmath $G$}^{-1} +
\left(
\begin{array}{ll}
U_{1}+ 2t(\cos k_x+ \cos k_y) & 0 \\
0     & U_{2}
\end{array}  \right) \right]^{-1}_{11}
\label{Gext}
\end{eqnarray} 

In summary if we treat bosons as a free Bose gas
and include
attraction induced by gauge and $|\chi_{ij}|$
fluctuations, the electron Green function is
approximately given by Eq.
\ref{Gext}.  However, different treatments of fluctuations result in
$\frac{U_{2}}{U_{1}}$ in a range
from $-1/3$ to 1 and absolute magnitudes of $U_{1,2}$ are of order $t$.

\begin{figure}
\centerline{\epsfig{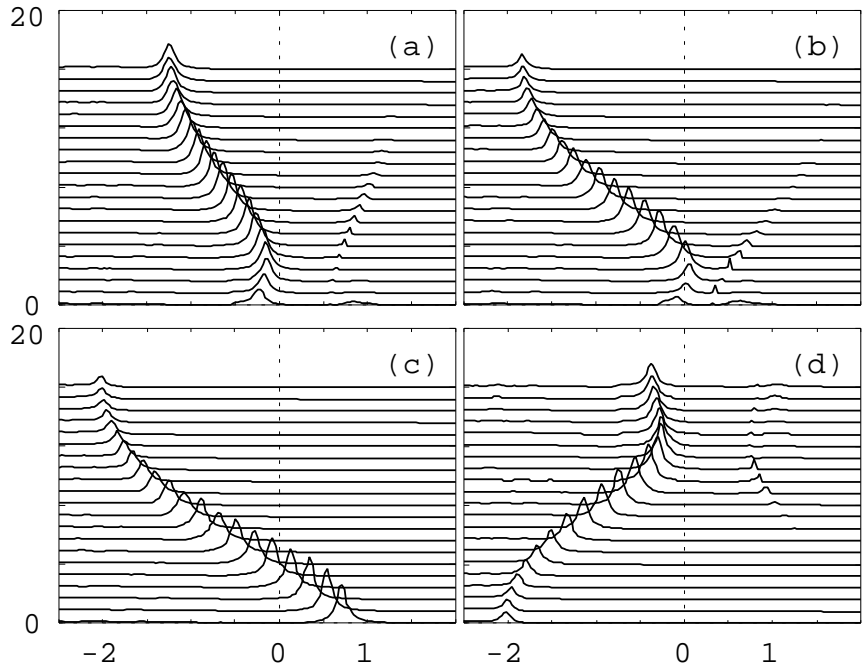}}
\caption{
The electron spectral function for $U_2/U_1=-1/3$ and,
from top down,
(a) $k=(-\pi/4, \pi/4) \to (\pi/4, 3\pi/4)$,
(b) $k=(-\pi/8, \pi/8) \to (3\pi/8, 5\pi/8)$,
(c) $k=(0, 0) \to (\pi/2, \pi/2)$, and
(d) $k=(0, \pi) \to (0,0)$.
We have chosen $J=1$.
}
\label{fig5}
\end{figure} 

\begin{figure}
\centerline{\epsfig{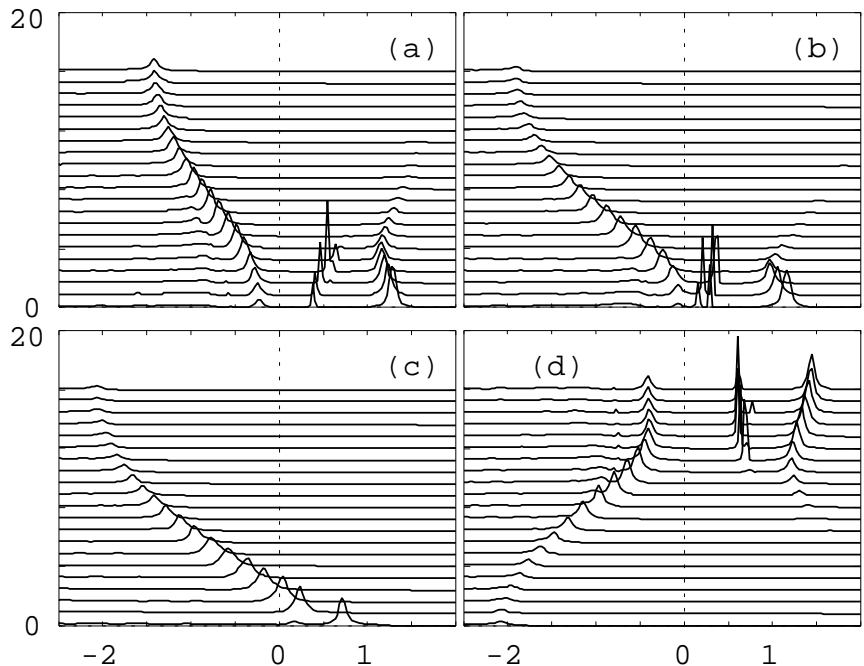}}
\caption{
The electron spectral function for $U_2/U_1=1$ and,
from top down
(a) $k=(-\pi/4, \pi/4) \to (\pi/4, 3\pi/4)$,
(b) $k=(-\pi/8, \pi/8) \to (3\pi/8, 5\pi/8)$,
(c) $k=(0, 0) \to (\pi/2, \pi/2)$, and
(d) $k=(0, \pi) \to (0,0)$.
We have chosen $J=1$.
}
\label{fig6}
\end{figure} 

In Fig. \ref{fig5} and \ref{fig6}
we plot the electron spectral function for
$\frac{U_{2}}{U_{1}} = -1/3, 1$.  We
have chosen $\tilde J=J/2$, $\tilde t=t=2J$,
$\chi = 1$, $\Delta/\chi = 0.4$, $x = 0.1$, and $T =
 0.1 J$.
The value $U_1$ is determined from requiring the renormalized
electron Green function to satisfies the sum rule
\begin{equation}
\int_0^\infty \frac{d\omega d^2k}{(2\pi)^3} \Im G_e= x
\end{equation} 

The main purpose of the above study is to understand the ambiguity in the
electron spectral
function due to our uncertainty in treating gauge fluctuation.  
We find
that the gap near $(0 \pm
\pi)$ and $(\pm \pi,0)$ survives the inclusion of gauge fluctuation.
However spectral functions
near $(\pm \frac{\pi}{2}, \pm \frac{\pi}{2})$ are modified.  
For $\omega<0$ the electron
spectral functions are quite
similar for the three different choices of 
$\frac{U_{2}}{U_{1}}=-1/3,\ 0,\ 1$.  But for
$\omega>0$ the spectral
functions show some notable differences.

\end{document}